\documentclass[twocolumn]{aastex631}

\usepackage{booktabs}
\usepackage{multirow}
\usepackage{float}
\usepackage{graphicx}	
\usepackage{amsmath}	
\usepackage{amssymb}	
\usepackage{breqn}
\usepackage[para,online,flushleft]{threeparttable}

\received{xxxx}
\revised{yyyy}
\accepted{zzzz}

\submitjournal{AJ}

\shorttitle{High-resolution spectroscopic analysis of four unevolved barium stars}
\shortauthors{Roriz et al.}

\begin{document}

\title{High-resolution spectroscopic analysis of four unevolved barium stars\footnote{HD~15096, HD~37792, and HD~141804 were
observed under the program ID 097.A-9024(A). HD~207585 was observed under the agreement between Observat\'orio Nacional (Brazil) and European Southern Observatory (ESO).}}

\correspondingauthor{M. P. Roriz}
\email{michelle@on.br}

\author[0000-0001-9164-2882]{M. P. Roriz}
\affiliation{Observat\'orio Nacional/MCTI, Rua General Jos\'e Cristino, 77, 20921-400, Rio de Janeiro, Brazil}

\author[0000-0002-8504-6248]{N. Holanda}
\affiliation{Observat\'orio Nacional/MCTI, Rua General Jos\'e Cristino, 77, 20921-400, Rio de Janeiro, Brazil}

\author[0000-0002-5042-443X ]{L. V. da Concei\c{c}\~ao}
\affiliation{Department of Physics and Astronomy, University of Manitoba, Winnipeg, MB, R3T 2N2, Canada}

\author{S. Junqueira}
\affiliation{Observat\'orio Nacional/MCTI, Rua General Jos\'e Cristino, 77, 20921-400, Rio de Janeiro, Brazil}

\author[0000-0003-4842-8834]{N. A. Drake}
\affiliation{Observat\'orio Nacional/MCTI, Rua General Jos\'e Cristino, 77, 20921-400, Rio de Janeiro, Brazil}
\affiliation{Laboratory of Observational Astrophysics, Saint Petersburg State University, Universitetski pr. 28, 198504, Saint Petersburg, Russia}

\author{A. Sonally}
\affiliation{Observat\'orio Nacional/MCTI, Rua General Jos\'e Cristino, 77, 20921-400, Rio de Janeiro, Brazil}

\author{C. B. Pereira}
\affiliation{Observat\'orio Nacional/MCTI, Rua General Jos\'e Cristino, 77, 20921-400, Rio de Janeiro, Brazil}



\begin{abstract}
A classical Local Thermodynamic Equilibrium analysis, based on high-resolution spectroscopic data, is performed for a sample of three potential barium dwarf candidates and one star already recognized as such. We derived their atmospheric parameters, estimated their masses and luminosities, and determined chemical abundances for a set of 21 elements, including CNO. Some elemental abundances are derived for the first time in HD~15096, HD~37792, and HD~141804. The program stars are dwarfs/subgiants with metallicities typical of disc stars, exhibiting moderate carbon enhancements, with $\rm{[C/Fe]}$ ratios ranging from $+0.29$ to $+0.66$ dex, and high levels of \textit{slow} neutron-capture ($s$-process) elements, with $\rm{[\textit{s}/Fe]}\gtrsim +1.0$~dex. As spectroscopic binaries, their peculiarities are attributable to mass transfer events. The observed neutron-capture patterns of were individually compared with two sets of $s$-process nucleosynthesis models (Monash and {\sc fruity}), yielding dilution factors and masses estimates for the former polluting Asymptotic Giant Branch stars. Low-mass ($\lesssim 3.0~\rm{M}_{\odot}$) models successfully reproduce the observations. In addition, we estimated mean neutron exposures of the order of 0.6 -- 0.7 mb$^{-1}$ for the $s$-processed material observed in their envelopes. Applying an empirical initial-final mass relation, we constraint in $\sim 0.7\,\textrm{M}_{\odot}$ the mass of their dim white-dwarf companions. Moreover, our kinematic study revealed that the program stars are members of the thin disk, with probabilities greater than 70\%. Hence, we identified HD~15096 and HD~37792 as new barium dwarfs and confirmed that HD~141804 is a barium dwarf. Thus, the number of barium dwarfs identified in the literature from high-resolution spectroscopy increases to 71 objects.
\end{abstract}




\section{Introduction}\label{sec:intro}

Barium (Ba) stars \citep[][]{bidelman1951}{}{} and their Population II analogues, CH stars \citep[][]{keenan1942}{}{}, were initially recognized as red giants enriched in carbon and elements synthesized mostly through the \textit{slow} neutron-capture mechanism \citep[$s$-process;][]{b2fh1957, kappeler2011, lugaro2023}{}{}. However, the $s$-process nucleosynthesis is expected to take place in the interiors of Thermally-Pulsing Asymptotic Giant Branch (TP-AGB) stars \citep[][]{gallino1998, busso1999, straniero2006, karakas2014}{}{}. Consequently, as first ascent giants, Ba and CH stars are not able to produce \textit{in loco} and self-enriched their envelopes with the $s$-processed material. Until the discovery of their binary nature \citep[][]{mcclure1980, mcclure1984, mcclure1990}{}{}{}, such peculiarities posed a challenge to the early stellar evolution models.

Belonging to binary systems, the chemical anomalies observed in Ba and related stars are attributed to mass exchange effects. In the framework of post-mass-transfer interacting binaries, the primary TP-AGB star loses mass and pollutes the atmosphere of its less evolved companion. As the secondary star evolves, it becomes a $s$-rich giant, observed as a Ba/CH star, depending on the metallicity, while the former TP-AGB ends as a dim White Dwarf (WD). Over the years, detailed chemical abundance analyses have confirmed the $s$-rich nature of the Ba stars \citep[][]{allen2006, pereira2011b, decastro2016, karinkuzhi2018b, shejeelammal2020, roriz2021a, roriz2021b}{}{} and CH stars \citep[][]{goswami2006, karinkuzhi2014, karinkuzhi2015, goswami2016, purandardas2019}{}{}. Moreover, their binary nature is widely supported by data acquired from extensive programs of radial velocity monitoring \citep[][]{jorissen1998, jorissen2019}. As post-mass-transfer binaries, Ba/CH stars provide valuables observational constraints to $s$-process nucleosynthesis models in AGB stars \citep[e.g.,][]{cseh2018, cseh2022}{}{}, binary star evolution models \citep[e.g.,][]{escorza2020}{}{}, and the mass-transfer mechanisms \citep[e.g.,][]{jorissen1998, jorissen2016}{}{}.

\subsection{Barium dwarf stars}

In addition to explaining the origins of the Ba/CH giants, the mass transfer scenario also predicts the existence of less evolved analogue stars exhibiting in their envelopes the Ba\,{\sc ii} syndrome observed in the classical giants. Indeed, the discovery of the so-called ``CH \textit{subgiant} stars'' \citep[][]{bond1974, luck1991, luck1982}{}{}{} and the F/G-type main-sequence Ba \textit{dwarf} stars \citep[][]{tomkin1989, north1994}, thought to be linked to the classical giants \citep[see, e.g.,][]{escorza2020}{}{}, provided observational evidence in this sense.

Despite their nomenclatures, Ba dwarfs and CH subgiants exhibit many similarities. They share the same region in the HR diagram \citep{escorza2017}, and exhibit no clear distinction in the period-eccentricity diagram \citep[][]{escorza2019, north2020}{}{}. From a chemical point of view, studies based on high-resolution spectroscopy point to the chemical similarity between Ba dwarfs and CH subgiants \citep[][]{pereira2003, pereira2005, allen2006}{}{}. These systems are generally referred to as Ba dwarfs and, just like the classical giants, also help us to trace back their former TP-AGB polluters, figuring as powerful tracers of the $s$-process nucleosynthesis.

\begin{figure}
    \centering
    \includegraphics{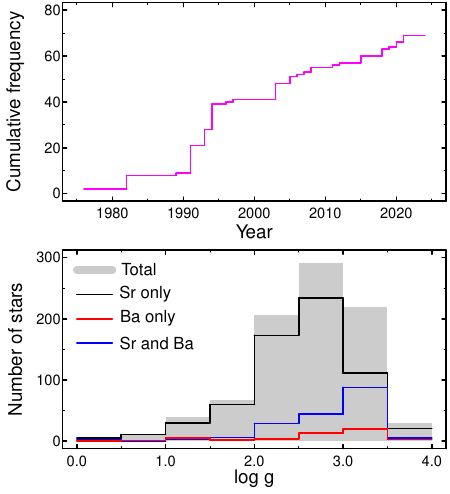}
    \caption{\textit{Upper panel}: Temporal evolution of the cumulative frequency of Ba dwarf stars confirmed from detailed chemical analyses. \textit{Bottom panel}: Distribution of the 895 $s$-process-rich candidates (grey) reported by \citet{norfolk2019}, identifying the sub-sample labeled as Sr only (black), Ba only (red), and Sr and Ba (blue). Stars with $\log g\geq 3.5$ represent only $\sim 0.3$~\% of the total number of the candidates.}
    \label{fig:cum_freq}
\end{figure}

However, Ba dwarfs turned out to be rare objects, calling into question whether they are the progenitors of the Ba/CH giants \citep[e.g.][]{luck1991}{}{}. In light of the mass transfer hypothesis, the former are expected to be as common as the latter \citep[][]{frantsman1992, han1995}{}{}. A search of the literature reveals that the current sample of Ba dwarfs, confirmed from detailed chemical analyses, comprises only 69 stars. This is a relatively small number, compared to the current sample of Ba giants studied from high-resolution spectroscopy; for example, \citet{decastro2016} reported an analysis of 169 Ba giants.

Moreover, as illustrated in the upper panel of Figure~\ref{fig:cum_freq}, the number of Ba dwarfs grown up very slowly over the years. In that plot, we show the temporal evolution of the cumulative frequency of these stars, starting from the first quantitative abundance analysis reported in the literature \citep[][]{sneden1976}{}{}. In recent years, Ba dwarfs has been found in the studies of \citet{kong2018}, \citet{purandardas2019}, \citet{shejeelammal2020}, \citet{liu2021}, and \citet{karinkuzhi2021}. We refer the reader to \citet[][and references therein]{kong2018} for a complete list until that year.

Recently, based on the Large Sky Area Multi-Object Fibre Spectroscopic Telescope ({\sc lamost}) data along with machine learning techniques, \citet{norfolk2019} were able to select a sample of 895 $s$-process-rich candidates, using absorption features of Ba\,{\sc ii} and Sr\,{\sc ii} in their spectra. These stars were classified into three groups: those with only Ba enhancement (Ba only), only Sr enhancement (Sr only), and both Ba and Sr enhancements (Ba and Sr). However, the Ba only index turned out to be more reliable proxy for identifying potential Ba star candidates \citep[see][]{karinkuzhi2021, guo2023}{}{}. We have found 29 stars with superficial gravities such that $\log g\geq3.5$ in the sample of \citeauthor{norfolk2019}; among them, only 3 stars are marked as Ba only, according the algorithm employed by \citeauthor{norfolk2019}. This represents a very low frequency of occurrence (only $\sim0.3$~\%). In the bottom panel of Figure~\ref{fig:cum_freq}, we show the distribution of the sample analyzed by \citeauthor{norfolk2019}. In other words, the task of finding Ba dwarfs is not easy, which reinforces the need of identifying new candidates and exploring their chemical patterns.

In the present work, we conduct a high-resolution spectroscopic analysis of two potential Ba dwarf candidates, HD\,15096 and HD\,37792. In addition to them, we have also include in our analysis the star HD\,141804, previously classified as CH subgiants by \citet{luck1991}, and a well-studied Ba dwarf (HD\,207585). In the following, we describe in Section~\ref{sec:observations} the criteria adopted in selection of the targets, as well as the details of the observation and data acquisition; in Section~\ref{sec:atm_par}, we give details of the spectroscopic analysis applied to the four stars; in Section~\ref{sec:abd_anal}, we describe the procedure to derive the elemental abundances; in Section~\ref{sec:discussion}, we discuss the abundance results in the literature context. In Section~\ref{sec:comparison}, we compare the observed patterns for the neutron-capture elements with predictions from the $s$-process nucleosynthesis models. In Section~\ref{sec:kinematic}, we carry out a kinematic study for the program stars. Concluding remarks are outlined in Section~\ref{sec:summary}.

\section{Program stars and Observations}\label{sec:observations}
\citet{fulbright2000} conducted a high-resolution spectroscopic analysis of 168 halo and disk stars, mainly dwarfs. Among them, we noticed that HD~15096 and HD~37792 stood out of the data set, showing relatively high barium abundances, with [Ba/Fe]\footnote{Throughout this paper, we use the standard spectroscopic notation, ${\rm [A/B]}=\log(N_{\rm A}/N_{\rm B})_{\star} -\log(N_{\rm A}/N_{\rm B})_{\odot}$, and the definition $\log\epsilon({\rm A})=\log(N_{\rm A}/N_{\rm H})+12$.} ratios of $+0.96$ and $+1.29$ dex, respectively. Such high values caught our attention to consider these two targets as chemically peculiar candidates. Additionally, we realized that the barium abundance of HD~15096 is also accompanied by similar values for the [Y/Fe] and [Zr/Fe] ratios. However, no other neutron-capture element abundance data were reported for HD~15096 and HD~37792 in the literature.

The stars HD~141804 and HD~207585, in turn, belong to Table 3 of \citet{luck1991}. That table lists 15 $s$-process enhanced stars with surface gravities ($\log g$) ranging from 3.0 to $\sim4.0$, so labeled as CH subgiants. Later works realized that some of these stars have different $\log g$ values, which put them in other evolutionary stage. This is the case of the stars BD$+$09$\degr$2384, CPD$-$62$\degr$6195 ($=$CD$-$62$\degr$1346), HD~122202, HD~123585, and HD~207585. For BD$+$09$\degr$2384 and CPD$-$62$\degr$6195, \citet{decastro2016} and \citet{pereira2012} found $\log g<3.0$, which evidenced the giant nature of these stars. For HD~122202, HD~123585, and HD~207585, investigations carried out by \citet{north1994}, \citet{allen2006}, \citet{karinkuzhi2015}, and \citet{shejeelammal2020} reported $\log g$ values greater than those reported by \citeauthor{luck1991}, thus evidencing the dwarf nature of these objects.

For HD~141804, there is no other analysis in addition to that performed by \citet{luck1991}. For this target, we present here for the first time elemental abundances of nitrogen, oxygen, aluminum, strontium, samarium, europium, and lead, as well as estimates of neutron exposure and a kinematic analysis. For HD~207585, on the other side, \citet{shejeelammal2020} provided detailed elemental abundances and recognized the Ba dwarf nature of this object. However, as we will demonstrate later, we have found significant differences regarding the nitrogen and oxygen abundances reported by these authors. Additionally, we derive for the first time strontium abundances and estimated the neutron exposure level for HD~207585.

The high-resolution spectra of the program stars were acquired using the Fiber-fed Extended Range Optical Spectrograph \citep[FEROS;][]{kaufer1999}{}{}, installed at the 2.2\,m Max Planck Gesellschaft (MPG)/European Southern Observatory (ESO) Telescope in La Silla, Chile. The observational missions were carried out between 2008 October and 2016 March. FEROS covers the spectral region between 3\,500 and 9\,200~\AA\, with a resolving power $R=\lambda$/$\Delta \lambda$\,$\approx$\,48\,000. In order to achieve a typical signal-noise ratio S/N~$\approx$\,150–200, the exposure time ranges from 600 to 1\,200\,s. We have used the FEROS Data Reduction System pipeline to reduce the observed spectra. General information about our targets are presented in Table~\ref{tab:general}, where we provided their positions, proper motions and parallaxes \citep{gaia2020}, V and B magnitudes \citep{zacharias2004}, observation dates, and corresponding exposure times.

\begin{table*}
\centering
\caption{General description of our sample.}\label{tab:general}
    \begin{tabular}{lccccccccc}
    \toprule
    Star  &  RA       &  DEC                &  V    &  B    &  pmRA            &  pmDEC           & Plx   & Date Obs.    & Exp. \\
          & (h\,m\,s) & ($^{\circ}$\,'\,'') & (mag) & (mag) & (mas\,yr$^{-1}$) & (mas\,yr$^{-1}$) & (mas) & (yyyy-mm-dd) &  (s)  \\ 
    \midrule
    HD\,15096  & 02\,26\,01.76 & $+$05\,46\,46.35 & 7.930 & 8.745  & $+$383.857 & $+$162.084 & 34.1998 & 2016-09-25 & 600 \\
    HD\,37792  & 05\,40\,17.43 & $-$19\,13\,37.77 & 7.722 & 8.093  & $-$127.539 & $-$192.658 & 13.3559 & 2016-09-09 & 600 \\
    HD\,141804 & 15\,53\,29.24 & $-$54\,09\,30.50 & 9.038 & 9.582  & $-$21.448  & $-$66.878  & 11.3044 & 2016-03-13 & 900 \\
    HD\,207585 & 21\,50\,34.71 & $-$24\,11\,11.69 & 9.785 & 10.482 & $+$14.390  & $-$36.805  & 4.8441  & 2008-10-19 & 1\,200 \\
    \bottomrule
    \end{tabular}
\end{table*}

\section{Atmospheric and physical parameters}\label{sec:atm_par}

The atmospheric parameters, effective temperature ($T_{\rm eff}$), $\log\,g$, microturbulent velocity ($\xi$), and metallicity ([Fe/H]), for the stars HD\,15096, HD\,37792, HD\,141804, and HD\,207585 were derived by applying the same procedure described, for example, in \citet[][]{roriz2017}, \citet{pereira2019b}, and \citet{holanda2020, holanda2023}. Firstly, we measured the equivalent widths (EW) of a set of Fe\,{\sc i} and Fe\,{\sc ii} absorption lines, by fitting Gaussian profiles to them. For this assignment, we have used the task {\sc splot} of the {\sc iraf} \citep[Image Reduction and Analysis Facility;][]{tody1986}. The atomic parameters of the Fe\,{\sc i} and Fe\,{\sc ii} transitions, such as excitation potential ($\chi$) and $\log$ \textit{gf} values, were taken from \citet{lambert1996}, i.e., the same line list consistently employed in the aforementioned references. The equivalent width measurements are listed in Table~\ref{app:ew_iron}. In the task of deriving the atmospheric parameters, we implemented the version 2013 of the {\sc moog}\footnote{Available at $\href{https://www.as.utexas.edu/~chris/moog.html}{https://www.as.utexas.edu/~chris/moog.html}$} spectral analysis code \citep{sneden1973, sneden2012}. {\sc moog} assumes the Local-Thermodynamic-Equilibrium (LTE) conditions and the plane-parallel atmosphere models. We have adopted the models computed by \citet{kurucz1993}. 

To find the effective temperature, we assumed the excitation equilibrium. Such a condition is verified when there is no trend between the iron abundances and the lower excitation potential of the measured Fe\,{\sc i} lines, or when the slope of the linear fit is very close to zero. We derived the microturbulent velocity by constraining the Fe\,{\sc i} lines until the iron abundance obtained from them shows no dependence with the reduced equivalent width ($\log {\rm EW}/{\lambda}$). The surface gravity is computed by imposing the ionization equilibrium, i.e., when the Fe\,{\sc i} and Fe\,{\sc ii} abundances are equal at the fixed $T_{\rm eff}$.

The errors associated with temperatures were estimated considering the uncertainty in the value of the slope of the linear fit for Fe\,{\sc i} abundances versus $\chi$. On the other hand, the errors associated with microturbulent velocities were estimated from the uncertainty in the slope of the linear fit for the same Fe\,{\sc i} abundances versus $\log {\rm EW}/{\lambda}$. The errors in surface gravities were estimated by changing the $\log g$ value until the difference in the average abundances of Fe\,{\sc i} and Fe\,{\sc ii} is equal to the standard deviation of the [Fe\,{\sc i}/H] mean. The final metallicity was normalized to the solar iron abundance recommendation of \citet{grevesse1998}, $\log\epsilon(\rm{Fe})=7.50$ dex. 

Table~\ref{tab:parameters} presents the atmospheric parameters derived in this work and their respective uncertainties, along with data compiled from the literature. In general, we found a good agreement between our results and those reported in previous studies. Furthermore, Table~\ref{tab:parameters} also lists the $\log g$ values derived from GAIA's parallaxes \citep[][]{gaia2020}{}{}; these data corroborate the consistency of our results. In particular, for HD~141804, we derived $\log g=4.50$, whereas \citet{luck1991} reported $\log g=3.50$, which leads us to conclude that HD~141804 is, in fact, a dwarf star. Regarding HD~207585, we remark the close agreement between the atmospheric parameters derived in this work with those reported recently by \citet{shejeelammal2020}. It is also worth noting that these $\log g$ values are slightly larger than the value derived by \citeauthor{luck1991}.

\begin{table*}
\centering
\caption{Adopted atmospheric and physical parameters for HD\,37792, HD\,15096, HD\,141804, and HD\,207585, in comparison with values previously reported in the literature.}\label{tab:parameters}
    \begin{tabular}{lcccccccc}
    \toprule
    Star   & $T_{\rm eff}$ & $\log\,g$      & [Fe/H] & $\xi$          & $\log\,g^{\rm GAIA}$ & Mass          & Reference \\  
           & (K)           & (cm\,s$^{-2}$) & (dex)  & (km\,s$^{-1}$) & (cm\,s$^{-2}$)       & (M$_{\odot}$) & \\ 
    \midrule
    HD~15096  & 5380$\pm$30   & 4.40$\pm$0.10 & $-$0.14$\pm$0.06 & 0.80$\pm$0.10 & 4.51$\pm$0.02 & 0.85$\pm$0.03 & This Work \\
           & 5375          & 4.30          & $-$0.20          & 0.80          & ---           & ---           & \citet{fulbright2000} \\
           & 5119          & 4.39          & $-$0.48          & 0.43          & ---           & ---           & \citet{gratton2003} \\
           & 5247          & 4.35          & $-$0.41          & ---           & ---           & ---           & \citet{soubiran2005} \\ 
    \midrule
    HD~37792  & 6500$\pm$20   & 4.10$\pm$0.20 & $-$0.55$\pm$0.09 & 1.50$\pm$0.10 & 4.09$\pm$0.03 & 1.07$\pm$0.04 & This Work \\
           & 6500          & 4.10          & $-$0.60          & 1.50          & ---           & ---           & \citet{fulbright2000} \\
    \midrule
    HD~141804 & 6230$\pm$50   & 4.50$\pm$0.10 & $-$0.41$\pm$0.08 & 1.20$\pm$0.10 & 4.35$\pm$0.03 & 0.99$\pm$0.04 & This Work \\
           & 6000          & 3.50          & $-$0.41          & 1.70          & ---           & ---           & \citet{luck1991} \\
    \midrule
    HD~207585 & 5840$\pm$50   & 3.90$\pm$0.20 & $-$0.34$\pm$0.08 & 1.10$\pm$0.10 & 3.84$\pm$0.03 & 1.10$\pm$0.04 & This Work \\
           & 5400          & 3.50          & $-$0.74          & ---           & ---           & ---           & \citet{smith1986a} \\
           & 5400$\pm$300  & 3.30$\pm$0.30 & $-$0.57$\pm$0.05 & 1.80$\pm$0.50 & ---           & ---           & \citet{luck1991} \\
           & 5800$\pm$100  & 3.80$\pm$0.20 & $-$0.38$\pm$0.12 & 1.00$\pm$0.20 & ---           & ---           & \citet{shejeelammal2020} \\
    \bottomrule
\end{tabular}
\end{table*}

\begin{figure}
    \centering
    \includegraphics{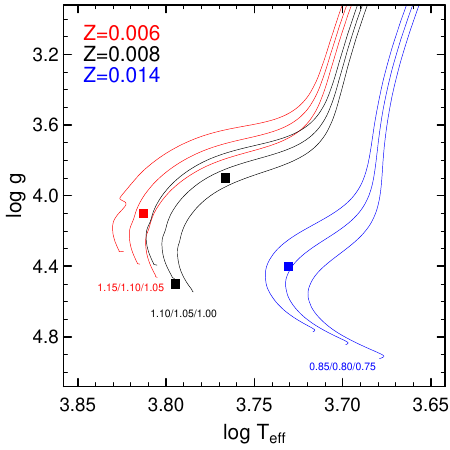}
    \caption{Position of the stars HD\,15096 (blue), HD\,37792 (red), HD\,141804 (black), and HD\,207585 (black) in the $\log\,T_{\rm eff}$ versus $\log\,g$ diagram. Evolutionary tracks from \citet{bressan2012}, for Z=0.006, 0.008, and 0.014, are also shown; the numbers correspond to stellar masses (in solar unit, $M_{\odot}$).}
    \label{fig:tracks}
\end{figure}

To estimate the masses and ages of our targets, we have used the {\sc parsec} (PAdova and TRieste Stellar Evolution Code; \citealp{bressan2012}) evolutionary tracks and a Bayesian estimation method \citep{dasilva2006}. This estimation has been performed through {\sc param}\footnote{Available online at $\href{http://stev.oapd.inaf.it/cgi-bin/param_1.3}{http://stev.oapd.inaf.it/cgi-bin/param_1.3}$}, a helpful tool to determine the basic intrinsic parameters of stars, given from their photometric and spectroscopic data. Individual values for masses and their respective uncertainties are also listed in Table~\ref{tab:parameters}. Additionally, the ages (in Gyr) derived from that procedure are 6.177$\pm$3.623 (HD~15096), 4.768$\pm$0.556 (HD~37792), 4.241$\pm$1.729 (HD~141804), and 5.858$\pm$0.561 (HD~207585). The positions of these stars in the Kiel Diagram are shown in Figure~\ref{fig:tracks}, along with evolutionary tracks for metallicities $Z$\,=\,0.006, 0.008, and 0.014. From this plot, we see that at least HD~207585 is in a post-main-sequence evolutionary stage.

Since effective temperatures, surface gravities, and stellar masses have been derived for the program stars, we can evaluate their luminosities from the relation:  

\begin{equation}
\log \left (\frac{L_{\star}}{L_{\odot}} \right )=4\log T_{\rm eff \star} - \log g_{\star} + \log \left (\frac{M_{\star}}{M_{\odot}} \right )-10.61,
\label{eq:lum_mass}
\end{equation}

\noindent for which we adopted $T_{\textrm{eff}\odot}=5\,777$ K and $\log g_{\odot}=4.44$. By inserting the data provided in Table~\ref{tab:parameters} into Equation~(\ref{eq:lum_mass}), we found $\log (L_{\star}/L_{\odot})=-0.16\pm0.10$, $+0.57\pm0.20$, $+0.06\pm0.10$, and $+0.60\pm0.20$, for HD~15096, HD~37792, HD~141804, and HD~207585, respectively. 

We can also derive their luminosities based on parallaxes, from the relation: 

\begin{equation}
\log \left (\frac{L_{\star}}{L_{\odot}} \right )=-2\log \pi + 0.4(M_{\rm{bol\odot}}-V+A_{V}-\rm{BC}-5),
\label{eq:lum_dist}
\end{equation}

\noindent where $\pi$ is the parallax (in arcsec), $M_{\rm{bol}\odot}=4.740\pm 0.024$ is the solar bolometric magnitude of \citet{bessel1998}, $V$ is the visual magnitude, BC is the bolometric correction, and $A_{V}$ is the  interstellar extinction. The bolometric corrections were evaluated from the empirical calibrations provided by \citet{alonso1995} for dwarf stars, whereas the $A_{V}$ values were estimated from the empirical extinction law of \citet{chen1998}. From this procedure, we have estimated $\log (L_{\star}/L_{\odot})=-0.31\pm0.04$, $+0.63\pm0.05$, $+0.28\pm0.06$, and $+0.65\pm0.05$, for HD~15096, HD~37792, HD~141804, and HD~207585, respectively.

\subsection{Radial velocity data}\label{sec:radial_velocity}

As will be discussed in Section~\ref{sec:discussion}, the chemical peculiarities observed in our stars cannot be explained in light of the stellar evolution of an isolated star, but in the mass transfer framework. Hence, in this section, we discuss the binary status of our targets. According to SIMBAD database, they are all classified as spectroscopic binaries. Three of them, HD~15096, HD~141804, and HD~207585, have their orbital elements already determined, and made available in the ninth catalogue of spectroscopic binary orbits (the SB9\footnote{Available on line at \href{https://sb9.astro.ulb.ac.be/}{https://sb9.astro.ulb.ac.be/}} database; \citealt{pourbaix2004}). HD~15096, HD~141804, and HD~207585 exhibit circular orbits (i.e., $e \sim 0.0$) with orbital periods (in days) of $3\,600\pm41$, $2\,652\pm95$, and $672\pm2$, respectively \citep[][]{latham2002, escorza2019}{}{}. As binary systems, their chemical peculiarities are attributable to mass transfer. We have also derived their radial velocities, from the Doppler shift of the spectral lines. We obtained, in km\,s$^{-1}$, $-5.625\pm0.194$, $+10.133\pm0.440$, $-55.301\pm0.406$, and $-55.891\pm0.309$ for HD~15096, HD~37792, HD~141804, and HD~207585, respectively.

\section{Abundance analysis}\label{sec:abd_anal}

\begin{table*}
    \centering
    \caption{Elemental abundances derived for the stars HD\,15096 and HD\,37792. For guidance, solar photospheric abundances adopted in this work \citep[][]{grevesse1998}{}{}{} are listed in the second column. Third and eighth columns provide the stellar abundances in the scale $\log\epsilon$({\rm H}) = 12.0; an asterisk is used to indicate abundances with non-LTE corrections. Abundances in the [X/H] and [X/Fe] notations are also shown. In columns 4 and 9, we provide the abundance dispersion ($\sigma_{\rm obs}$) due to line-to-line scatter. These were evaluated when three or more transitions are considered, except for strontium and elements whose abundances were derived from spectral synthesis (syn), as explained in the text. For abundances based on equivalent width measurements, we provide information on the number of spectral lines used. At the end, we give the mean carbon abundance, considering the abundances derived from atomic absorption lines and the synthesis of molecular bands with the respective dispersion. In this case $n$(\#) is the the number of the abundances obtained.} \label{tab:abd_hd37792_hd15096}
        \begin{tabular}{@{\extracolsep{4pt}}lccccccccccccc}
        \toprule
                & & & \multicolumn{5}{c}{HD\,15096} & & \multicolumn{5}{c}{HD\,37792}\\
                \cline{4-8}
                \cline{10-14}
        Species & $\log\epsilon_{\odot}$ & & $\log\epsilon$ &  $\sigma_{\rm obs}$ & $n$(\#) & [X/H] & [X/Fe]  & & $\log\epsilon$ & $\sigma_{\rm obs}$ & $n$(\#) & [X/H] & [X/Fe] \\
        \midrule 
        C\,{\sc i}           & 8.52 & & 8.67 & 0.09 & 03   & $+$0.15 & $+$0.29 & & 8.43 & 0.10 & 11   & $-$0.09 & $+$0.46 \\
        C\,(C$_{2}$ 5\,165)  & 8.52 & & 8.70 & 0.02 & syn  & $+$0.18 & $+$0.32 & & ---  & ---  & ---  & ---     & ---     \\
        C\,(C$_{2}$ 5\,635)  & 8.52 & & 8.65 & 0.03 & syn  & $+$0.13 & $+$0.27 & & ---  & ---  & ---  & ---     & ---     \\
        N                    & 7.92 & & 8.07 & 0.10 & syn  & $+$0.15 & $+$0.29 & & ---  & ---  & ---  & ---     & ---     \\
        O\,{\sc i}           & 8.83 & & 8.83 & 0.33 & syn  &    0.00 & $+$0.14 & & 8.55$^{*}$ & 0.04 & 03   & $-$0.28 & $+$0.27 \\
        Na\,{\sc i}          & 6.33 & & 6.18 & 0.06 & 04   & $-$0.15 & $-$0.01 & & 5.97 & 0.05 & 04   & $-$0.36 & $+$0.19 \\
        Mg\,{\sc i}          & 7.58 & & 7.58 & 0.10 & 05   &    0.00 & $+$0.14 & & 7.18 & 0.10 & 07   & $-$0.40 & $+$0.15 \\
        Al\,{\sc i}          & 6.47 & & 6.24 & 0.05 & 06   & $-$0.23 & $-$0.09 & & ---  &  --- & ---  &     --- &     --- \\
        Si\,{\sc i}          & 7.55 & & 7.51 & 0.04 & 05   & $-$0.04 & $+$0.10 & & 7.17 & 0.07 & 05   & $-$0.38 & $+$0.17 \\
        Ca\,{\sc i}          & 6.36 & & 6.22 & 0.08 & 10   & $-$0.14 &    0.00 & & 5.93 & 0.09 & 15   & $-$0.43 & $+$0.12 \\
        Ti\,{\sc i}          & 5.02 & & 4.79 & 0.09 & 20   & $-$0.23 & $-$0.09 & & 4.42 & 0.08 & 14   & $-$0.60 & $-$0.05 \\
        Cr\,{\sc i}          & 5.67 & & 5.44 & 0.06 & 05   & $-$0.23 & $-$0.09 & & 5.02 & 0.07 & 08   & $-$0.65 & $-$0.10 \\
        Fe\,{\sc i}          & 7.50 & & 7.36 & 0.06 & 89   & $-$0.14 &    ---  & & 6.95 & 0.09 & 74   & $-$0.55 &     --- \\
        Fe\,{\sc ii}         & 7.50 & & 7.35 & 0.05 & 14   & $-$0.15 &    ---  & & 6.95 & 0.07 & 12   & $-$0.55 &     --- \\
        Ni\,{\sc i}          & 6.25 & & 6.05 & 0.08 & 21   & $-$0.20 & $-$0.06 & & 5.85 & 0.11 & 07   & $-$0.40 & $+$0.15 \\
        Sr\,{\sc i}          & 2.97 & & 3.70 & 0.04 & 01   & $+$0.73 & $+$0.87 & & 3.04 & 0.07 & 01   & $+$0.07 & $+$0.62 \\
        Y\,{\sc ii}          & 2.24 & & 3.07 & 0.07 & 05   & $+$0.83 & $+$0.97 & & 2.54 & 0.09 & 06   & $+$0.30 & $+$0.85 \\
        Zr\,{\sc ii}         & 2.60 & & 3.51 & 0.14 & 04   & $+$0.91 & $+$1.05 & & 2.98 & 0.11 & 04   & $+$0.38 & $+$0.93 \\
        Ba\,{\sc ii}         & 2.13 & & 3.13 & 0.06 & syn  & $+$1.00 & $+$1.14 & & 3.03 & 0.10 & syn  & $+$0.90 & $+$1.45 \\
        La\,{\sc ii}         & 1.17 & & 1.96 & 0.11 & 05   & $+$0.79 & $+$0.93 & & 1.59 & 0.09 & 03   & $+$0.42 & $+$0.97 \\
        Ce\,{\sc ii}         & 1.58 & & 2.47 & 0.10 & 07   & $+$0.89 & $+$1.03 & & 2.05 & 0.08 & 06   & $+$0.47 & $+$1.02 \\
        Nd\,{\sc ii}         & 1.50 & & 2.25 & 0.05 & 07   & $+$0.75 & $+$0.89 & & 1.93 & 0.11 & 07   & $+$0.43 & $+$0.98 \\ 
        Sm\,{\sc ii}         & 1.01 & & 1.68 & 0.03 & 04   & $+$0.67 & $+$0.81 & & 1.78 & ---  & 02   & $+$0.77 & $+$1.32 \\ 
        Eu\,{\sc ii}         & 0.51 & & 0.71 & 0.09 & syn  & $+$0.20 & $+$0.34 & & ---  & ---  & ---  & ---     & ---     \\
        Pb\,{\sc i}          & 1.95 & & 2.85 & 0.10 & syn  & $+$0.90 & $+$1.04 & & $<$1.90 & ---  & syn  & $<-$0.05 & $<+$0.50 \\        
        \midrule
        C                    & 8.52 & & 8.67 & 0.03 & 03   & $+$0.15 & $+$0.29 & & ---  & ---  & ---  & ---     & ---     \\
        \bottomrule
    \end{tabular}
\end{table*}

\begin{table*}
    \centering
    \caption{The same as in Table~\ref{tab:abd_hd37792_hd15096}, for the stars HD\,141804 and HD\,207585.}\label{tab:abd_hd141804_hd207585}
        \begin{tabular}{@{\extracolsep{4pt}}lccccccccccccc}
        \toprule
                & & & \multicolumn{5}{c}{HD\,141804} & & \multicolumn{5}{c}{HD\,207585}\\
                \cline{4-8}
                \cline{10-14}
        Species & $\log\epsilon_{\odot}$ & & $\log\epsilon$ &  $\sigma_{\rm obs}$ & $n$(\#) & [X/H] & [X/Fe]  & & $\log\epsilon$ & $\sigma_{\rm obs}$ & $n$(\#) & [X/H] & [X/Fe] \\
        \midrule
        C\,{\sc i}           & 8.52 & & 8.85 & 0.08 & 09  & $+$0.33 & $+$0.74 & & 8.97 & 0.09  & 09  & $+$0.45 & $+$0.79 \\  
        C\,(C$_{2}$ 5\,165)  & 8.52 & & 8.72 & 0.03 & syn & $+$0.20 & $+$0.61 & & 8.72 & 0.03  & syn & $+$0.20 & $+$0.54 \\
        C\,(C$_{2}$ 5\,635)  & 8.52 & & 8.62 & 0.10 & syn & $+$0.10 & $+$0.51 & & 8.82 & 0.03  & syn & $+$0.30 & $+$0.64 \\
        N                    & 7.92 & & 7.62 & 0.03 & syn & $-$0.30 & $+$0.11 & & 7.52 & 0.04  & syn & $-$0.40 & $-$0.06 \\
        O\,{\sc i}           & 8.83 & & 8.58$^{*}$ & 0.04 & 03  & $-$0.25 & $+$0.16 & & 8.55$^{*}$ & 0.10  & 03  & $-$0.28 & $+$0.06 \\
        Na\,{\sc i}          & 6.33 & & 6.06 & 0.06 & 04  & $-$0.27 & $+$0.14 & & 6.18 & 0.10  & 04  & $-$0.15 & $+$0.19 \\
        Mg\,{\sc i}          & 7.58 & & 7.29 & 0.12 & 04  & $-$0.29 & $+$0.12 & & 7.33 & 0.06  & 03  & $-$0.25 & $+$0.09 \\
        Al\,{\sc i}          & 6.47 & & 5.96 & 0.07 & 03  & $-$0.51 & $-$0.10 & & 5.96 & 0.07  & 04  & $-$0.51 & $-$0.17 \\
        Si\,{\sc i}          & 7.55 & & 7.32 & 0.04 & 03  & $-$0.23 & $+$0.18 & & 7.39 & 0.07  & 03  & $-$0.16 & $+$0.18 \\
        Ca\,{\sc i}          & 6.36 & & 6.01 & 0.07 & 16  & $-$0.35 & $+$0.06 & & 6.09 & 0.09  & 17  & $-$0.27 & $+$0.07 \\
        Ti\,{\sc i}          & 5.02 & & 4.54 & 0.06 & 16  & $-$0.48 & $-$0.07 & & 4.56 & 0.08  & 18  & $-$0.46 & $-$0.12 \\
        Cr\,{\sc i}          & 5.67 & & 5.22 & 0.07 & 08  & $-$0.45 & $-$0.04 & & 5.30 & 0.10  & 05  & $-$0.37 & $-$0.03 \\
        Fe\,{\sc i}          & 7.50 & & 7.09 & 0.08 & 64  & $-$0.41 &     --- & & 7.16 & 0.08  & 73  & $-$0.34 &    ---  \\
        Fe\,{\sc ii}         & 7.50 & & 7.10 & 0.07 & 12  & $-$0.40 &     --- & & 7.16 & 0.08  & 11  & $-$0.34 &    ---  \\
        Ni\,{\sc i}          & 6.25 & & 5.92 & 0.08 & 10  & $-$0.33 & $+$0.08 & & 5.96 & 0.09  & 09  & $-$0.29 & $+$0.05 \\
        Sr\,{\sc i}          & 2.97 & & 3.81 & 0.07 & 01  & $+$0.84 & $+$1.25 & & 3.88 & 0.07  & 01  & $+$0.91 & $+$1.25 \\
        Y\,{\sc ii}          & 2.24 & & 3.18 & 0.05 & 06  & $+$0.94 & $+$1.35 & & 3.24 & 0.10  & 06  & $+$1.00 & $+$1.34 \\
        Zr\,{\sc ii}         & 2.60 & & 3.67 & 0.11 & 07  & $+$1.07 & $+$1.48 & & 3.78 & 0.11  & 06  & $+$1.18 & $+$1.52 \\
        Ba\,{\sc ii}         & 2.13 & & 3.43 & 0.06 & syn & $+$1.30 & $+$1.71 & & 3.53 & 0.02  & syn & $+$1.40 & $+$1.74 \\
        La\,{\sc ii}         & 1.17 & & 2.29 & 0.05 & 05  & $+$1.12 & $+$1.53 & & 2.21 & 0.09  & 05  & $+$1.04 & $+$1.38 \\
        Ce\,{\sc ii}         & 1.58 & & 2.86 & 0.09 & 12  & $+$1.28 & $+$1.69 & & 2.84 & 0.09  & 06  & $+$1.26 & $+$1.60 \\
        Nd\,{\sc ii}         & 1.50 & & 2.68 & 0.08 & 14  & $+$1.18 & $+$1.59 & & 2.61 & 0.06  & 10  & $+$1.11 & $+$1.45 \\
        Sm\,{\sc ii}         & 1.01 & & 1.94 & 0.07 & 06  & $+$0.93 & $+$1.34 & & 1.88 & 0.07  & 07  & $+$0.87 & $+$1.21 \\   
        Eu\,{\sc ii}         & 0.51 & & 0.88 & 0.05 & syn & $+$0.37 & $+$0.78 & & 0.76 & 0.04  & syn & $+$0.25 & $+$0.59 \\
        Pb\,{\sc i}          & 1.95 & & 3.00 & 0.13 & syn & $+$1.05 & $+$1.46 & & 2.96 & 0.10  & syn & $+$1.01 & $+$1.35 \\
        \midrule
        C                    & 8.52 & & 8.73 & 0.12 & 03  & $+$0.21 & $+$0.62 & & 8.84 & 0.13  & 03  & $+$0.32 & $+$0.66 \\
        \bottomrule
    \end{tabular}
\end{table*}

\begin{table*}
\centering
\caption[]{Comparison between elemental abundances derived in (1) this work, (2) \citet{fulbright2000}, (3) \citet{luck1991}, (4) \citet{shejeelammal2020}, and (5) \citet{masseron2010}.}
\label{tab:literature}
	\begin{tabular}{lccccccccccc}
    \toprule
    Star       & [C/Fe] & [N/Fe] & [O/Fe] & [Na/Fe] & [Mg/Fe] & [Al/Fe] & [Si/Fe] & [Ca/Fe] & [Ti/Fe] & [Cr/Fe] & Ref.  \\
    \midrule
    HD\,15096  & $+$0.29 & $+$0.29 & $+$0.14 & $-$0.01 & $+$0.14 & $-$0.09 & $+$0.10 &    0.00 & $-$0.09 & $-$0.09 &  1 \\
               &   --    &   --    &   --    & $+$0.04 & $+$0.21 & $+$0.19 & $+$0.08 & $+$0.09 & $+$0.13 & $-$0.01 &  2 \\[0.3cm]
    HD\,37792  & $+$0.46 &   --    & $+$0.27 & $+$0.19 & $+$0.15 &   --    & $+$0.17 & $+$0.17 & $-$0.05 & $-$0.10 &  1 \\
               &   --    &   --    &   --    & $+$0.13 & $+$0.11 &   --    & $+$0.31 & $+$0.15 & $+$0.33 &   --    &  2 \\[0.3cm]
    HD~141804  & $+$0.62 & $+$0.11 & $+$0.16 & $+$0.14 & $+$0.12 & $-$0.10 & $+$0.18 & $+$0.06 & $-$0.07 & $-$0.04 &  1 \\
               & $+$0.24 &   --    &   --    & $+$0.01 & $-$0.08 &   --    & $+$0.14 & $+$0.15 & $+$0.32 & $-$0.03 &  3 \\[0.3cm]
    HD\,207585 & $+$0.66 & $-$0.06 & $+$0.06 & $+$0.19 & $+$0.09 & $-$0.17 & $+$0.18 & $+$0.07 & $-$0.12 & $-$0.03 &  1 \\
               &   --    &   --    &   --    & $+$0.25 & $+$0.02 &   --    & $+$0.28 & $+$0.39 & $+$0.01 & $+$0.24 &  3 \\
               & $+$0.61 & $+$0.75 & $+$0.97 & $+$0.25 & $+$0.07 &   --    & $+$0.12 & $+$0.26 & $+$0.01 & $+$0.13 &  4 \\
               & $+$0.51 & $+$0.12 & $+$0.14 &   --    & $-$0.03 &   --    &   --    &   --    &   --    &   --    &  5 \\
    \midrule
    Star       & [Ni/Fe] & [Y/Fe]  & [Zr/Fe] & [Ba/Fe] & [La/Fe] & [Ce/Fe] & [Nd/Fe] & [Sm/Fe] & [Eu/Fe] & [Pb/Fe] & Ref. \\
    \midrule
    HD\,15096  & $-$0.06 & $+$0.97 & $+$1.05 & $+$1.14 & $+$0.93 & $+$1.03 & $+$0.89 & $+$0.81 & $+$0.34 & $+$1.04 &  1 \\
               &    0.00 & $+$1.02 & $+$1.00 & $+$0.96 &   --    &   --    &   --    &   --    & $+$0.30 &   --    &  2 \\[0.3cm]
    HD\,37792  & $+$0.15 & $+$0.85 & $+$0.93 & $+$1.45 & $+$0.97 & $+$1.02 & $+$0.98 & $+$1.32 &   --    & $+$0.50 &  1 \\
               & $+$0.10 &   --    &   --    & $+$1.29 &   --    &   --    &   --    &   --    &   --    &   --    &  2 \\[0.3cm]
    HD~141804  & $+$0.08 & $+$1.35 & $+$1.48 & $+$1.71 & $+$1.53 & $+$1.69 & $+$1.59 & $+$1.34 & $+$0.78 & $+$1.46 &  1 \\
               & $+$0.02 & $+$0.60 & $+$1.48 & $+$1.25 & $+$1.18 & $+$1.10 & $+$0.95 &   --    &  --     &   --    &  3 \\[0.3cm]
    HD\,207585 & $+$0.05 & $+$1.34 & $+$1.52 & $+$1.74 & $+$1.38 & $+$1.60 & $+$1.45 & $+$1.21 & $+$0.59 & $+$1.35 &  1 \\
               & $+$0.08 & $+$1.30 & $+$1.02 &   --    & $+$1.61 & $+$0.85 & $+$0.94 & $+$1.06 &   --    &   --    &  2 \\
               & $-$0.01 & $+$1.37 & $+$1.20 & $+$1.60 & $+$1.70 & $+$1.72 & $+$1.62 & $+$2.04 & $+$0.28 &   --    &  3 \\
               &   --    &   --    &   --    & $+$1.23 & $+$1.37 & $+$1.41 &   --    &   --    & $+$0.58 & $+$1.30 &  4 \\
    \bottomrule
	\end{tabular}
\end{table*}

Chemical abundances were derived for a set of 21 different elements, based on either equivalent width measurements or synthetic spectra computations of selected absorption lines. To carry out this task, we have used the drivers \textit{abfind}, \textit{blends}, and \textit{synth} of {\sc moog}. The final abundances are presented in Tables~\ref{tab:abd_hd37792_hd15096} and \ref{tab:abd_hd141804_hd207585}, and a comparison with literature values is provided in Table~\ref{tab:literature}. Our abundance data are normalized to the solar value recommended by \citet{grevesse1998}.

\begin{figure}
    \centering
    \includegraphics[width=\columnwidth]{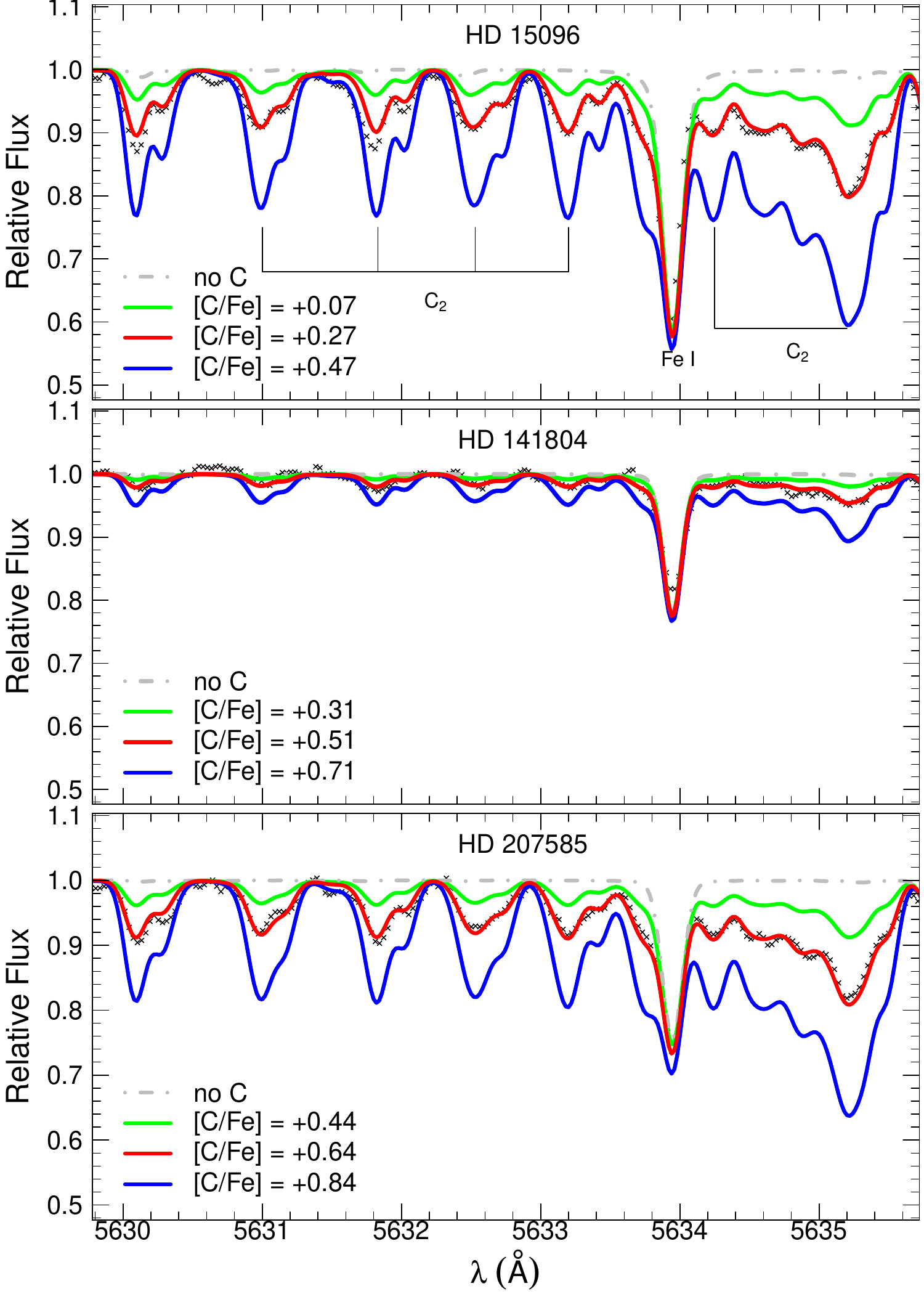}
    \caption{Observed (crosses) and synthetic (lines) spectra around the spectral region of the C$_{2}$ molecular band at 5\,635~\AA\, for the stars HD\,15096, HD\,141804, and HD\,207585. The synthetic spectra were computed for different [C/Fe] values, as indicated in each panel. The gray lines are spectral synthesis without contribution of the C$_{2}$ molecule.}
    \label{fig:syn_c2_5635}
\end{figure}

Carbon abundances were determined from equivalent width measurements of some absorption atomic lines of C\,{\sc i} and also by computing synthetic spectra of molecular band heads for the stars HD~15096, HD~141804, and HD~207585. For HD~37792, the warmer (6\,500~K) star of the program, the carbon molecular features were present for carbon abundance determination. Therefore, carbon abundance for this object was estimated only based on the C\,{\sc i} atomic lines present in its spectrum, as seen in Table~\ref{tab:abd_hd37792_hd15096}. To synthesized the molecular carbon features observed in HD~15096, HD~141804, and HD~207585, we used the C$_{2}$ (0,0) band head of the Swan system $d^3\Pi_{g} - a^3\Pi_{u}$ at 5\,165~\AA, as performed in \citet{roriz2017}, and the C$_{2}$ (0,1) band head of the Swan system $d^3\Pi_{g} - a^3\Pi_{u}$ at 5\,635~\AA, as in \citet{drake2008}. Figure~\ref{fig:syn_c2_5635} shows the observed and synthetic spectra around the spectral region of the C$_{2}$ molecular band at 5\,635~\AA, for these stars. 

Nitrogen abundances were also obtained via spectral synthesis technique. For HD~15096, we used the $^{12}$CN molecular lines of the system $A^2\Pi - X^2\Sigma$ in the 7\,994--8\,020~\AA\ wavelength range, adopting the same linelist provided in \citet{drake2008}. For the stars HD~141804 and HD~207585, we used the $B^2\Sigma - X^2\Sigma$ violet system band head at 3\,883~\AA, as done in \citet{roriz2017}, with the linelist provided by the Vienna Atomic Line Database ({\sc vald}; \citealt{ryabchikova2015}). For HD~37792, however, we were not able to derive its nitrogen abundance. Additionally, the carbon isotopic ratios, $^{12}$C/$^{13}$C, could not be measured in our program stars, as they are too hot for identify the $^{13}$CN lines. It is also worth mentioning that nitrogen abundances in Ba dwarfs are reported in the literature for a very limited number of stars \citep[][]{reddy2003, allen2006, drake2007, purandardas2019, shejeelammal2020, liu2021}{}{}. Among them, many values are upper limits for the [N/Fe] ratios.

Concerning oxygen abundances, we have used the [O\,{\sc i}] forbidden line at 6\,300.3~\AA\ for the star HD~15096, and adopted $\log$ \textit{gf} $=-9.72$ from \citet{allende2001}. For HD~37792, HD~141804, and HD~207585, oxygen abundances were evaluated from the equivalent width measurements of the oxygen infrared triplet around 7\,774~\AA. Taking into account the well known non-LTE effects in the oxygen infrared triplet lines, we performed non-LTE corrections for abundances derived from the O\,{\sc i} triplet lines. For this, we used yields of \citet{amarsi2015}, which are available by the {\sc inspect} database, version 1.0 (\href{http://www.inspect-stars.com/}{http://www.inspect-stars.com/}). After the non-LTE corrections, the [O/Fe] ratios for HD~141804 and HD~207585 were lowered by 0.22 and 0.31 dex, respectively, while for HD~15096 the corrected [O/Fe] ratio increased in 0.02 dex. The adopted values are marked with an asterisk in Tables~\ref{tab:abd_hd37792_hd15096} and \ref{tab:abd_hd141804_hd207585}.

For the light elements Na, Mg, Al, Si, Ca, and Ti, iron-group elements Cr and Ni, and the neutron-capture elements Sr, Y, Zr, La, Ce, Nd, and Sm, abundances were derived from equivalent width measurements of selected atomic lines sufficiently unblended to provide reliable abundances of such species. In Table~\ref{app:ew_other} of Appendix, we provide the adopted lab data, such as wavelength, excitation potentials, and $\log$ \textit{gf} values, of the atomic transitions used in our analysis, as well as the measured equivalent widths. For the element lanthanum, in particular, we have followed the same procedure employed by \citet{roriz2021b}. In other words, we measured the equivalent widths of the lanthanum lines and run driver {\sl blends} of {\sc moog}, which provides abundances from blended spectral lines, to take into account the hyperfine splitting (HFS) that strongly affect the transitions of this element. 

Barium, europium, and lead abundances, on the other hand, were determined from synthetic spectrum analysis. For barium abundances, we used the Ba\,{\sc ii} line at 5\,853.7~\AA, and adopted its HFS data from \citet{mcwilliam1998}. For europium, we considered the Eu\,{\sc ii} line at 6\,645.1~\AA, for which HFS components were taken from \citet{lawler2001b}. Lead abundances were derived from the Pb\,{\sc i} line at 4\,057.8~\AA. HFS and isotopic data for lead were taken from \citet{vaneck2003}.

\subsection{Abundance uncertainties}

In the task of deriving chemical abundances, two uncertainty sources should be taken into account: ($i$) the line parameters, as equivalent width measurements, oscillator strength values, continuum normalization, and line blending, which introduce random errors ($\sigma_{\textrm{ran}}$) in abundances, as well as ($ii$) the errors associated with stellar parameters of atmospheric models. For a generic element X, the total uncertainty in $\log \epsilon (\textrm{X})$ can be evaluated according to the following equation:

\begin{dmath}\label{eq:uncertainty}
     \sigma_{\log \epsilon(\textrm{X})_{\star}}^{2}=\sigma_{\textrm{ran}}^{2}+\left(\frac{\partial \log \epsilon}{\partial T_{\textrm{eff}}}\right)^{2}\sigma_{T_{\textrm{eff}}}^{2}+ \left(\frac{\partial \log \epsilon}{\partial \log g}\right)^{2}\sigma_{\log g}^{2}+ \left(\frac{\partial \log \epsilon}{\partial \xi}\right)^{2}\sigma_{\xi}^{2}+ \left(\frac{\partial \log \epsilon}{\partial \textrm{[Fe/H]}}\right)^{2}\sigma_{\textrm{[Fe/H]}}^{2}+ \left(\frac{\partial \log \epsilon}{\partial W_{\lambda}}\right)^{2}\sigma_{W_{\lambda}}^{2}.
\end{dmath}

\noindent To evaluate the partial derivatives of the above equation, we shifted the parameters $T_{\rm{eff}}$, $\log g$, $\xi$, [Fe/H], and $W_{\lambda}$ in $+30$~K, $+0.1$~dex, $+0.1$~km\,s$^{-1}$, $+0.1$ dex, and $+3$~m\AA, respectively, which are typical uncertainties. Then, we computed the corresponding change introduced in the abundance when we vary one of the parameters, keeping the others fixed. For a resolution $R=48\,000$ and a typical signal-to-noise ratio S/N $\sim 100-200$, the expected uncertainties in the equivalent widths are of the order of $2-3$ m\AA, according to \citet{cayrel1988}. The $\sigma_{\textrm{ran}}$ term in equation~(\ref{eq:uncertainty}) takes into account the line-to-line scatter in abundance. It is evaluated as the ratio $\sigma_{\textrm{obs}}/\sqrt{n}$, where $\sigma_{\textrm{obs}}$ is the standard deviation and $n$ is the number of spectral lines considered in the abundance derivation. Thus, the uncertainty in the [X/Fe] ratios is given by:

\begin{equation}
    \sigma_{\textrm{[X/Fe]}}^{2}=\sigma_{\textrm{X}}^{2}+\sigma_{\textrm{Fe}}^{2}.
\end{equation}

The partial derivatives in Equation~(\ref{eq:uncertainty}) were evaluated for HD~15096, taken as a template star. Table~\ref{tab:uncertainties} shows as the abundances for HD~15096 change in response to changes in $T_{\rm{eff}}$, $\log g$, $\xi$, [Fe/H], and $W_{\lambda}$; these values were assumed for the other targets. On the other hand, $\sigma_{\textrm{ran}}$ was computed for each object, when three or more lines are used to derive abundances, except for strontium and elements whose abundances were derived from spectral synthesis. For these particular cases, $\sigma_{\rm obs}$ is evaluated from three different positions of the continuum.

For CNO abundances, the uncertainties were evaluated similarly. By varying one of the parameters $T_{\textrm{eff}}$, $\log g$, and $\xi$, keeping the others two constant, we computed the respective changes introduced in abundances of HD~15096. Additionally, since CNO abundances are not independent of each other, we also evaluated the variations introduced as a consequence of changes in $\log \epsilon(\rm{C})$, $\log \epsilon(\rm{N})$, and $\log \epsilon (\rm{O})$ individually. The results for the CNO uncertainty estimates are presented in Table~\ref{tab:uncertainties_cno}.

\begin{table*}
    \centering
    \caption{Abundance uncertainties for HD\,15096, taken as a template star. Second column gives the random errors, given by $\sigma_{\rm{ran}}$\,=\,$\sigma_{\rm obs}$/$\sqrt{n}$, where $n$ is the number of absorption lines used for the abundance determination. Columns from 3 to 7 show variations in the abundances introduced by changes in $T_{\rm eff}$, $\log g$, $\xi$, [Fe/H], and equivalent width measurements ($W_\lambda$), respectively. By combining quadratically the terms from the 2nd to 7th, we estimate the total uncertainties, listed in column 8. The last column provides the abundance dispersion in abundance due to line-to-line scatter, previously shown in Table~\ref{tab:abd_hd37792_hd15096}.}\label{tab:uncertainties}
        \begin{tabular}{lcccccccc}
        \toprule
        Species                     & 
        $\sigma_{\rm{ran}}$ & 
        $\Delta T_{\rm eff}$        & 
        $\Delta\log g$              & 
        $\Delta\xi$                 & 
        $\Delta$[Fe/H]              & 
        $\Delta W_{\lambda}$        & 
        $\sqrt{\sum \sigma^2}$      & 
        $\sigma_{\rm obs}$         \\
                       &      & ($+$30\,K) & ($+$0.1 dex)  & ($+$0.1 km\,s$^{-1}$) & ($-$0.1 dex) & ($+$3 m\AA) &  \\
      \midrule
         C\,{\sc i}    & 0.05 & $-$0.02  & $+$0.01 &    0.00 &    0.00 & $+$0.08 &  0.10 & 0.09 (03)\\
        Na\,{\sc i}    & 0.03 & $+$0.02  & $-$0.02 & $-$0.01 & $-$0.01 & $+$0.04 &  0.06 & 0.06 (04) \\         
        Mg\,{\sc i}    & 0.04 & $+$0.02  & $-$0.01 & $-$0.01 & $-$0.01 & $+$0.04 &  0.06 & 0.10 (05) \\         
        Al\,{\sc i}    & 0.02 & $+$0.02  & $-$0.01 &    0.00 &    0.00 & $+$0.05 &  0.06 & 0.05 (06) \\         
        Si\,{\sc i}    & 0.02 & $-$0.01  &    0.00 & $-$0.01 & $-$0.02 & $+$0.05 &  0.06 & 0.04 (05) \\         
        Ca\,{\sc i}    & 0.03 & $+$0.03  & $-$0.02 & $-$0.02 & $-$0.01 & $+$0.05 &  0.07 & 0.08 (10) \\         
        Ti\,{\sc i}    & 0.03 & $+$0.04  &    0.00 & $-$0.02 &    0.00 & $+$0.07 &  0.09 & 0.09 (20) \\ 
        Cr\,{\sc i}    & 0.03 & $+$0.04  & $-$0.01 & $-$0.03 & $-$0.01 & $+$0.06 &  0.08 & 0.06 (05) \\ 
        Fe\,{\sc i}    & 0.01 & $+$0.01  & $-$0.01 & $-$0.03 & $-$0.02 & $+$0.06 &  0.07 & 0.06 (89) \\ 
        Fe\,{\sc ii}   & 0.01 & $-$0.02  & $+$0.03 & $-$0.02 & $-$0.03 & $+$0.08 &  0.10 & 0.05 (14) \\       
        Ni\,{\sc i}    & 0.02 & $+$0.01  & $+$0.01 & $-$0.02 & $-$0.02 & $+$0.07 &  0.08 & 0.08 (21) \\         
        Sr\,{\sc i}    & 0.02 & $+$0.04  & $-$0.03 & $-$0.03 & $-$0.03 & $+$0.04 &  0.08 & 0.04 (01) \\         
         Y\,{\sc ii}   & 0.03 & $+$0.01  & $+$0.02 & $-$0.04 & $-$0.03 & $+$0.07 &  0.09 & 0.07 (05) \\         
        Zr\,{\sc ii}   & 0.07 &    0.00  & $+$0.03 & $-$0.04 & $-$0.03 & $+$0.09 &  0.13 & 0.14 (04) \\
        Ba\,{\sc ii}   & 0.03 &    0.00  &    0.00 & $-$0.10 & $-$0.10 &    ---  &  0.14 & 0.06 (syn)  \\         
        La\,{\sc ii}   & 0.05 & $-$0.01  & $+$0.02 & $-$0.01 & $-$0.05 & $+$0.07 &  0.10 & 0.11 (05) \\         
        Ce\,{\sc ii}   & 0.04 & $+$0.01  & $+$0.04 & $-$0.03 & $-$0.03 & $+$0.10 &  0.12 & 0.10 (07) \\         
        Nd\,{\sc ii}   & 0.02 & $+$0.01  & $+$0.04 & $-$0.02 & $-$0.04 & $+$0.11 &  0.13 & 0.05 (07) \\           
        Sm\,{\sc ii}   & 0.02 & $+$0.01  & $+$0.04 & $-$0.02 & $-$0.03 & $+$0.11 &  0.12 & 0.03 (04) \\         
        Eu\,{\sc ii}   & 0.05 & $-$0.05  &    0.00 &    0.00 & $-$0.05 &    ---  &  0.09 & 0.09 (syn)  \\
        Pb\,{\sc i}    & 0.06 &    0.00  &    0.00 & $-$0.10 & $-$0.10 &    ---  &  0.15 & 0.10 (syn)  \\
        \bottomrule
    \end{tabular}
\end{table*}

\begin{table*}
    \centering
    \caption{Abundance uncertainties of carbon, nitrogen, and oxygen for the star HD\,15096.}\label{tab:uncertainties_cno} 
        \begin{tabular}{ccccccccc}
        \toprule
        Species                      & 
        $\Delta T_{\rm eff}$         & 
        $\Delta\log g$ & $\Delta\xi$ & 
        $\Delta\log {\rm (C)}$       & 
        $\Delta\log{\rm (N)}$        & 
        $\Delta\log{\rm (O)}$        & 
        $\sqrt{\sum \sigma^2}$           \\
           &   ($+$30~K) &   ($+$0.1 dex)   & ($+$0.1~km\,s$^{-1}$)  & ($+$0.20 dex)  & ($+$0.20 dex)   & ($+$0.20 dex)   &       \\
        \midrule
        C  &  $-$0.01  &  $+$0.02   &  $-$0.01  &     ---  &   0.00    &  $+$0.07  &  0.07 \\
        N  &     0.00  &     0.00   &  $-$0.02  &  $-$0.20 &   ---     &  $-$0.14  &  0.24 \\
        O  &     0.00  &  $+$0.05   &  $+$0.05  &  $-$0.02 &   0.00    &     ---   &  0.07 \\
        \bottomrule
    \end{tabular}
\end{table*}

\section{Results and Discussion}\label{sec:discussion}

\subsection{Carbon, nitrogen, and oxygen}\label{sub:sub:cno}
Elemental abundances of carbon, nitrogen, and oxygen are sensitive indexes to stellar evolution stages, as well as the mixing events that take place within the stars and alter the chemical composition of their atmospheres. According to stellar evolution models, when a star becomes a giant, the nuclear material previously exposed to CN-cycle is brought to the stellar surface, as a result of convective motions, known as first dredge-up (FDU), changing the initial abundance pattern of the star. As an outcome of FDU, carbon is depleted and nitrogen is increased on the stellar surface, while the oxygen content remains unchanged. Unlike the classical Ba/CH stars, Ba dwarfs did not experience the FDU. Consequently, the $s$-processed material received from their companions is unmixed with the nuclear material internally processed in the star. This is particularly useful for studying carbon and nitrogen, whose abundances should reflect the added material. 

\begin{figure}
    \centering
    \includegraphics{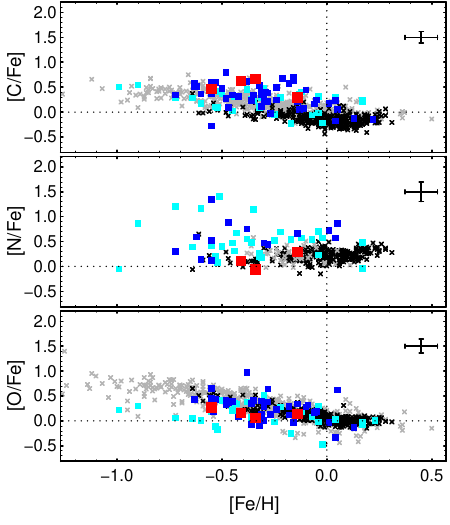}
    \caption{Abundance ratios for the elements carbon, nitrogen, and oxygen. The observed [C, N, O/Fe] ratios for the program stars (red squares) are plotted as a function of metallicity. Typical error bars are shown in the top of the panels. Data for field dwarf stars (gray crosses), field giant stars (black crosses), Ba giant stars (cyan squares), and Ba dwarf/CH subgiant stars (blue squares), collected from different literature sources, are also plotted. Data for field giant stars were taken from \citet{luck2007}; data for field dwarf stars were taken from \citet{reddy2003}, \citet{reddy2006}, and \citet{luck2006}; data for Ba giant stars were taken from \citet{allen2006}, \citet{decastro2016}, \citet{karinkuzhi2018b}, \citet{shejeelammal2020}, and \citet{roriz2021b}; data for Ba dwarfs/CH subgiant stars were taken from \citet{smith1993}, \citet{edvardsson1993}, \citet{north1994}, \citet{porto1997}, \citet{reddy2003}, \citet{pereira2003}, \citet{pereira2005}, \citet{allen2006}, \citet{pereira2011a}, \citet{karinkuzhi2015}, \citet{kong2018}, \citet{purandardas2019}, \citet{shejeelammal2020}, \citet{liu2021}, and \citet{karinkuzhi2021}.}
    \label{fig:cno}
\end{figure}

In Figure~\ref{fig:cno}, we show the [C, N, O/Fe] ratios observed for HD~15096, HD~37792, HD~141804, and HD~207585 as function of metallicity. We have also added in this figure data compiled from different sources of the literature for dwarf and giant normal stars, classical Ba giants, and Ba dwarf stars (i.e., with $\log g \ge 3.5$). The program stars exhibit moderate levels of carbon enhancements, with [C/Fe] ratios ranging from $\sim +0.30$ to $\sim +0.70$ dex, consistent with values already reported in the literature for Ba dwarfs of same metallicity. 

Regarding nitrogen abundances in Ba dwarfs, the scarcity of data in the literature is noteworthy. Among the reported data, most of them are upper limits \citep[see][]{allen2006}{}{}, and are not plotted in Figure~\ref{fig:cno}. It is remarked that Ba stars present generally a scatter larger than the observed in normal giant and dwarf stars, and some of them exhibit high levels of N abundances. It is known that massive ($\gtrsim 4.0~\rm{M}_{\odot}$) TP-AGB stars are able to produce high amounts of nitrogen through the Hot Bottom Burning (HBB) phenomenon \citep[][]{boothroyd1995}, when the convective envelope of the AGB star deepens its base into the H-burning shell. In principle, this mechanism could account the nitrogen excess observed in some Ba stars, however, the patterns observed for heavy elements in Ba stars evidence the low-mass ($\lesssim 3.0~\rm{M}_{\odot}$) nature of the polluting AGB stars \citep[e.g.,][]{cseh2018, karinkuzhi2018b, shejeelammal2020, roriz2021a, roriz2021b, cseh2022}.

For HD~15096, HD~141804, and HD~207585, we have derived $\textrm{[N/Fe]}=+0.29$, $+0.11$, and $-0.06$ dex, respectively. These values are commonly found in normal stars, as seen in the middle panel of Figure~\ref{fig:comparison}. This in turn could indicate inefficiencies in the CN-cycle and/or FDU of the ancient polluting stars, when they became giants. For HD~37792, our warmest (6\,500 K) star, N abundance could not be determined. As we mentioned in Section~\ref{sec:observations}, HD~207585 is a common star with the sample of \citet{shejeelammal2020}, who reported $\textrm{[N/Fe]}=+0.75$ dex (see Table~\ref{tab:literature}), while we have found a significantly lower value ($\textrm{[N/Fe]}=-0.06$ dex). For the same object, \citet{masseron2010} reported $\textrm{[N/Fe]}=+0.12$ dex. Similar disagreement concerns to oxygen abundances of HD~207585. In this work, we found $\textrm{[N/Fe]}=+0.06$ dex and \citet{masseron2010} $\textrm{[N/Fe]}=+0.14$ dex, while \citeauthor{shejeelammal2020} reported a much higher value, $\textrm{[O/Fe]}=+0.97$ dex.

For oxygen, the third panel of Figure~\ref{fig:cno} shows that our stars follow the Galactic trend, as expected. Additionally, we have derived their respective C/O ratios, which are commonly used to constraint Ba stars ($\rm{C/O}<1$; \citealt{barbuy1992}; \citealt{allen2006}; \citealt{pereira2009}; \citealt{karinkuzhi2018a, karinkuzhi2018b}; \citealt{roriz2023}) and CH stars ($\rm{C/O}>1$; \citealt{pereira2009}; \citealt{pereira2012}; \citealt{goswami2016}; \citealt{purandardas2019}). For HD~15096, HD~37792, HD~141804, and HD~207585, we found $\rm{C/O}=0.69, 0.76, 1.41$, and $1.95$, respectively.

\begin{figure}
    \centering
    \includegraphics{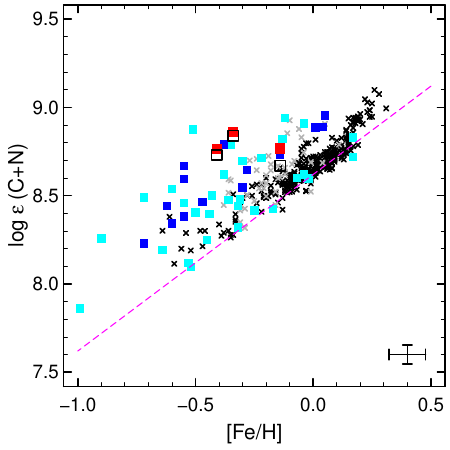}
    \caption{Combined carbon plus nitrogen abundances in the notation $\log\epsilon(\rm{C}+\rm{N})$ observed in HD~15096, HD~141804, and HD~207585 as a function of the metallicity. A typical error bar is also shown. The symbols have the same meaning as in Figure~\ref{fig:cno}. To illustrate the role of carbon in the CNO excess in our stars, we show the respective logarithmic carbon abundances in empty black squares. The magenta dashed line shows the initial CN abundance for a given metallicity.}
    \label{fig:log_cn}
\end{figure}

Considering the CNO abundance data, we can infer that carbon is actually the main responsible for the CNO excess observed in HD~15096, HD~141804, and HD~207585. This is also seen from their C+N combined abundances, for which we have $\log \epsilon(\rm{C}+\rm{N})=8.76, 8.76,$ and $8.86$, respectively. In Figure~\ref{fig:log_cn}, we compare the observed values of $\log \epsilon(\rm{C}+\rm{N})$ with those found in Ba dwarfs, Ba giants, and normal stars. It is notable that Ba stars present in this plane a scatter greater than the observed in normal field stars. Our stars lie close to other Ba dwarfs, which exhibit values systematically larger than those typically found in Ba giants. In conclusion, we see that the atmospheres of our stars were, indeed, contaminated by material previously exposed to He-burning shell.

\subsection{Elements from sodium to nickel}

We extracted chemical abundances for Na, $\alpha$-elements (Mg, Al, Si, Ca, and Ti), and iron-group (Cr and Ni) elements. Nucleosynthesis computations performed by \citet{woosley1995} predict that the elements sodium, magnesium, aluminum, silicon, calcium, and titanium in the Galaxy are mainly produced in hydrostatic burning environments of stars with initial masses of $10-40~\rm{M}_{\odot}$, as well as in supernova events. Sodium, magnesium, and aluminum are by-products of carbon burning, while most silicon, calcium, and titanium are produced from oxygen burning. Magnesium nucleosynthesis has also a contribution from hydrostatic neon burning. Type II supernova events are able to synthesize silicon and calcium, whereas the elements titanium, iron, chromium, and nickel, belonging to iron-group, are mainly produced from Type Ia supernovae. Therefore, Ba stars are expected to follow the Galactic trend for these elements. According to data in Tables~\ref{tab:abd_hd37792_hd15096} and \ref{tab:abd_hd141804_hd207585}, the [X/Fe] ratios for elements from Na to Ni are close to zero, similar to values observed in normal field stars.

In particular, for the odd-Z element Na, the [Na/Fe] ratios observed in field stars of the Galaxy show no trend for $\rm{[Fe/H]}\gtrsim -1.0$ dex \citep[e.g.][]{luck2006, luck2007}{}{}. For the studied stars, we found the [Na/Fe] ratios ranging from $-0.03$ to $+0.19$ dex, consistent with values reported for Ba dwarfs \citep[e.g.][]{allen2006}{}{}. Na can be produced through the NeNa-chain, during the H-burning in the convective core of main-sequence stars with mass $M\gtrsim 1.5~\rm{M}_{\odot}$. Then, when the star becomes a giant, FDU brings Na to the stellar surface, along with the by-products of CNO-cycle. However, since our targets did not reach the giant-branch stage of their evolution, their internally processed nuclear material was not dredged yet, so that our targets behave as normal stars. Works in the literature \citep[][]{antipova2004, decastro2016, karinkuzhi2018b, shejeelammal2020}{}{} reported $\rm{[Na/Fe]}\gtrsim +0.4$ dex for Ba giants, and \citet{decastro2016} observed an anti-correlation between [Na/Fe] and $\log g$ for the stars of their sample (see their Figure~20).

Na production is also associated with nucleosynthesis in AGB stars. Models predict that AGB stars are able to produce $^{23}$Na from Ne isotopes, through the NeNa-chain of proton capture, via $^{22}$Ne(\rm{p},$\gamma$)$^{23}$Na reaction, where the $^{22}$Ne was previously synthesized in the He-burning shell, from the $^{14}$N($\alpha$,$\gamma$)$^{18}$F($\beta^{+}$)$^{18}$O($\alpha$,$\gamma$)$^{22}$Ne reaction sequences \citep[e.g.,][]{mowlavi1999, karakas2014}. The processed material is mixed and brought to the AGB atmosphere, via Third Dredge-Up (TDU), and then is subsequently transferred to the observed Ba star. However no Na enhancement is observed in our stars.

\subsection{Neutron-capture elements}\label{sub:sub:heavy_elements}

Elements beyond the iron-peak ($Z>30$) are primarily synthesized throughout neutron-captures on seed nuclei, the $s$-process and the so-called $r$- (\textit{rapid}) process, depending on the time involved between neutron captures and $\beta$ decays. While the $s$-process takes place within TP-AGB stars, the exact astrophysical site(s) of the $r$-process is(are) a matter of debate, although supernovae events and neutron star mergers figure among the most promising scenarios \citep[see the review of][]{cowan2021}. As far as heavy elements are concerned, we have derived for our sample elemental abundances for Sr, Y, Zr, Ba, La, Ce, Nd, Sm, Eu, and Pb. As anticipated in Introduction, the chemical patterns of heavy elements observed in Ba stars help us to trace back their former polluter TP-AGB stars. 

\begin{table}
\centering
\caption{Contribution (\%) of the $s$-process to the solar-system material as provided by \citet[][A99]{arlandini1999} and \citet[][B14]{bisterzo2014}.} \label{tab:s-fraction}
        \begin{tabular}{lcc}
        \toprule

        Species &   A99                 & B14 \\
	    \midrule
        Sr      &   85                  &  69   \\
        Y       &   92                  &  72   \\
        Zr      &   83                  &  66   \\
        Ba      &   81                  &  85   \\
        La      &   62                  &  76   \\
        Ce      &   77                  &  84   \\
        Nd      &   56                  &  58   \\
        Sm      &   29                  &  31   \\
        Eu      &   6                   &  6    \\
        Pb      &   46                  &  87   \\
    	\bottomrule
     
        \end{tabular}

\end{table}

Table~\ref{tab:s-fraction} lists the $s$-process contribution to the Solar system material, where we can see a high contribution of the $s$-process in nucleosynthesis of the elements Sr, Y, Zr, Ba, La, Ce, and Nd. Indeed, they are found very abundant in our targets, with [X/Fe] ratios $\gtrsim +1.00$ dex, evidencing the strong $s$-rich nature of these objects, while normal field stars show [X/Fe] close to zero. Lead abundances are also enhanced in HD~15096, HD~141804, and HD~207585, with $\textrm{[Pb/Fe]}=+1.04$, $+1.46$, and $+1.35$ dex, respectively. For HD~37792, we were able to report only the upper limit $\textrm{[Pb/Fe]}<+0.50$ dex. Data on lead abundances, however, are very scarce in the literature. For the element samarium, which has a low $s$-process contribution (see Table~\ref{tab:s-fraction}), we also found excesses in our stars ($+0.80<\rm{[Sm/Fe]}<+1.35$). These high levels are not effects of the Galactic chemical evolution, and are found in Ba giants and other Ba dwarfs \citep[][]{allen2006}{}{}. Europium, a representative element of the $r$-process, shows slightly enhanced abundances in our stars, with [Eu/Fe] values of $+0.34$, $+0.78$, and $+0.59$ dex for HD~15096, HD~141804, and HD~207585, respectively. For HD~37792, europium abundance could not be derived, since the diagnostic spectral line was faintly detectable in its spectrum. 

\begin{figure}
    \centering
    \includegraphics{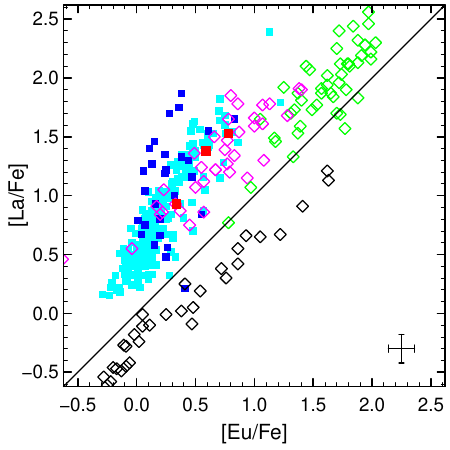}
    \caption{Program stars in the [La/Fe] versus [Eu/Fe] plane, along with data for other Ba dwarfs and Ba giants. Symbols and the literature references are the same listed in the caption of Figure~\ref{fig:cno}. Diamonds are data for CEMP-$r$ (black), CEMP-$s$ (magenta), and CEMP-$r/s$ (green), taken from \citet{masseron2010} and \citet{karinkuzhi2021}.}
    \label{fig:la_eu}
\end{figure}

In Figure~\ref{fig:la_eu}, we located our stars in the [La/Fe] versus [Eu/Fe] diagram, along with data available for other Ba dwarfs. We also added to this plot data for Ba giants and CEMP (Carbon-Enhanced Metal-Poor) stars \citep[][]{beers2005, masseron2010}{}{}. Since La and Eu are representative elements of the $s$- and $r$-processes, respectively, this diagram is able to split the data according the neutron-capture levels of enrichment observed in the stars. As demonstrated in that figure, our stars share the same portion (the $s$-rich side) where other Ba dwarfs, Ba giants, and the CEMP-$s$ lie.

\subsubsection{The s-process indexes}

\begin{figure}
    \centering
    \includegraphics{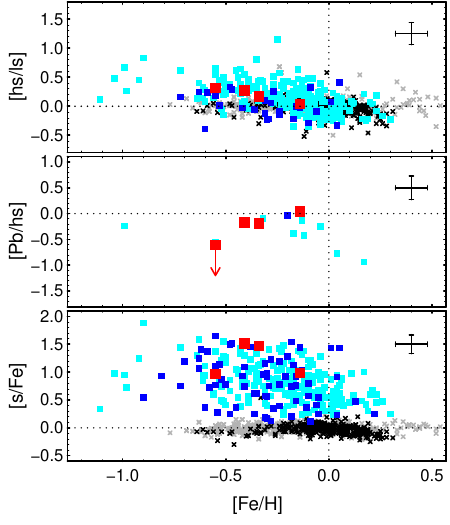}
    \caption{Indexes of the $s$-process. \textit{Upper}, \textit{middle}, and \textit{bottom} panels show respectively the [hs/ls], [Pb/hs], and [$s$/Fe] ratios as function of metallicity. Symbols and the literature references are the same listed in the caption of Figure~\ref{fig:cno}. Typical error bars for our data are also showed in the panels.}
    \label{fig:s-index}
\end{figure}

The $\rm{[hs/ls]}$ ratio is an intrinsic $s$-process index, widely used to probe the neutron exposures of the $s$-process \citep[e.g.,][]{luck1991, busso2001}{}{}. This ratio is defined as the difference $\rm{[hs/Fe]}-\rm{[ls/Fe]}$, where [hs/Fe] and [ls/Fe] are the averaged abundances of the elements belonging to the second (Ba, La, Ce, Nd) and first (Sr, Y, Zr) $s$-process peaks. High neutron exposures push nucleosynthesis of the second peak elements, instead elements of the first peak, so that [hs/ls]$>0$ is expected. HD~15096, HD~37792, HD~141804, and HD~207585, show $\rm{[hs/ls]}=+0.04$, $+0.31$, $+0.27$, and $+0.17$ dex, respectively. As shown in the upper panel of Figure~\ref{fig:s-index}, these values are consistent those reported in the literature for Ba stars. The [Pb/hs] ratio is useful to probe the nucleosynthesis among the third (Pb) and second peaks. We found $\rm{[Pb/hs]}=+0.04$, $-0.61$, $-0.17$, and $-0.19$ dex for HD~15096, HD~37792, HD~141804, and HD~207585, respectively. However, the literature lacks for lead abundance data (see middle panel of Figure~\ref{fig:s-index}). Considering the [hs/ls] and [Pb/hs] ratios, we conclude that the neutron exposures favored the nucleosynthesis of the second peak elements. 

We also computed the [$s$/Fe] index, given by the $s$-process average abundances from elemental abundances for Sr, Y, Zr, Ba, La, Ce, Nd, and Pb, yielding $\rm{[\textit{s}/Fe]}=+0.99$, $+0.92$, $+1.10$, and $+1.45$ dex for HD~15096, HD~37792, HD~141804, and HD~207585, respectively. Normal field stars typically show $\rm{[\textit{s}/Fe]}\sim 0$, whereas Ba stars exhibit [$s$/Fe] ratios increasing for lower metallicity regimes (see the bottom panel of Figure~\ref{fig:s-index}), a feature of the $s$-process \citep[e.g.][]{busso2001}{}{}. In their study on Ba giant stars, \citet{decastro2016} assumed [$s$/Fe]$>+0.25$ dex as a criterion to classify a star as Ba star, and ours stars satisfy this condition.

\subsubsection{Mean neutron exposures}

To infer quantitatively on the neutron exposure levels to which the $s$-processed material observed in our targets was subjected, we have used the classical analysis of the main $s$-process component \citep[][]{kappeler1990, kappeler2011}{}{} to estimate the mean neutron exposures ($\tau_{0}$). This approach consists of the so-called $\sigma N_{s}$ curve, which show the behavior of the product between the neutron-capture cross-sections ($\sigma$) and the abundances of the main $s$-process component ($N_{s}$), traditionally expressed in the scale $\log\epsilon(\textrm{Si})=6.0$. The $\sigma N_{s}$ curve is written in terms of $\tau_{0}$ and $G$, which are free parameters to fit the observations, where $G$ is the fraction of $^{56}$Fe used as seed of the $s$-process. $G$ has the effect of shifting the curve, while $\tau_0$ changes the shape of the $\sigma N_{s}$ curve. An interesting feature of the classical approach is that it successfully reproduces the main component of the $s$-process of the solar system, with $\tau_{0}\sim0.30$ mb$^{-1}$ \citep[][]{arlandini1999}{}{}, even without any assumption of the astrophysical site of the $s$-process.

\begin{figure}
    \centering
    \includegraphics{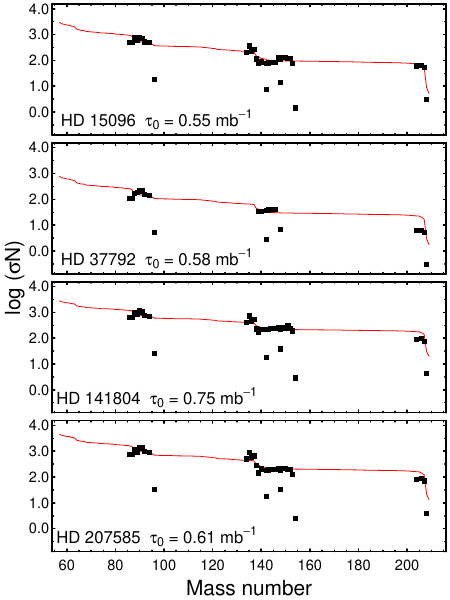}
    \caption{Fitting of theoretical $\sigma N_{s}$ curves (red lines) and observed $\sigma N_{s}$ values (black squares). The data are plotted in logarithmic scale; $\sigma$ is expressed in mb and $N_{s}$ in the scale $\log\epsilon(\textrm{Si})=6.0$. The estimated neutron exposures are shown in the bottom of panels.}
    \label{fig:sigman}
\end{figure}

From their abundance data in Ba stars, \citet{allen2006b} extracted abundances owing the main component of the $s$-process, and were able to estimate the neutron exposures for the stars of their sample, fitting $\sigma N_{s}$ curves to the observed data. For a detailed description of that procedure, we refer the reader to the respective paper. Here, we have applied a similar procedure to evaluate $\tau_{0}$ in our targets. In order to test our algorithm, we downloaded from the KADoNiS\footnote{Karlsruhe Astrophysical Database of Nucleosynthesis in Stars; available at \url{https://www.kadonis.org/}.} database a list of neutron-capture cross-sections (at 30 keV) and computed, via $\chi^{2}$ minimization, the best $\sigma N_{s}$ curve that reproduces the main component of the $s$-process for the solar distribution reported by \citet{arlandini1999}. With our implementation, we found $\tau_{0}=0.36$ mb$^{-1}$, close to value of \citeauthor{arlandini1999}; in their methodology, \citeauthor{allen2006b} found $\tau_{0}\sim0.35$ mb$^{-1}$. Extending that approach for HD~15096, HD~37792, HD~141804, and HD~207585, we found $\tau_{0}=0.55$, 0.58, 0.75, and 0.61 mb$^{-1}$, respectively. Figure~\ref{fig:sigman} shows the best fits of the $\sigma N_{s}$ curves to the observed data. In a recent review, \citet{lugaro2023} comment that the $^{13}$C neutron source (see next section) provides neutron exposure values up to $1.0$~mb$^{-1}$, which validates the results derived from our simple approach. \citeauthor{allen2006b} reported $\tau_{0}$ values ranging from 0.187 to 1.05 mb$^{-1}$ for the Ba stars of their sample.

\section{Comparison with s-process nucleosynthesis models}\label{sec:comparison}

Various groups have computed theoretical $s$-process yields from their AGB stellar models. In particular, the models provided by the INAF group \citep[][]{cristallo2009, cristallo2011, cristallo2015} -- through the on-line {\sc fruity}\footnote{FUll-Network Repository of Updated Isotopic Tables \& Yields, at \url{http://fruity.oa-teramo.inaf.it/}.} databse --  and Monash group \citep[][]{fishlock2014, karakas2016, karakas2018} comprise predictions for a wide range of masses ($1.0\leq M/\rm{M}_{\odot}\leq 8.0$) and metallicities ($-$1.2 $\lesssim$ [Fe/H] $\lesssim +0.3$), which covers the interval observed in Ba stars. 

TP-AGB stars provide a conducive environment for the $s$-process nucleosynthesis within a tenuous He-rich region (He-intershell) located between the H- and He-burning shells, which are activated alternately in this evolutionary stage \citep[][]{busso1999, straniero2006, kappeler2011, karakas2014, lugaro2023}{}{}. For low-mass ($1-3$ M$_{\odot}$) AGB stars, the $^{13}$C$(\alpha,\rm{n})^{16}$O reaction provides the main supply of neutrons to feed the $s$-process, during the interpulse periods (H-burning shell). This reaction is efficiently activated at $T\sim10^{8}$~K, releasing within the He-intershell a relatively low neutron density, of the order of $10^{7-8}$~cm$^{-3}$ \citep[][]{straniero1995, straniero1997}{}{}. The H-burning ashes accumulate in the inner layers and trigger periodically He-flashes (thermal pulses; TP) that, in turn, expand the upper layers of the stars. As a consequence, the H-burning in a shell is temporarily extinguished and a convective zone is driven in the He-intershell. Temperatures in this region can reach values high enough to activate an alternative neutron source, the $^{22}$Ne$(\alpha,\rm{n})^{25}$Mg reaction. However, for low-mass AGB stars, this reaction is only marginally activated. After a limited number of TP episodes, the base of the extensive convective envelope can deepen its base, carrying to the stellar surface (Third Dredge-Up; TDU) the by-products of the internal nucleosynthesis, rich in carbon and $s$-elements. The $^{22}$Ne$(\alpha,\rm{n})^{25}$Mg reaction becomes the main neutron provider for AGB stars of intermediate masses ($4-8$ M$_{\odot}$), during the TP events, at temperatures higher than $3\times10^{8}$~K, releasing a neutron burst of $10^{10-12}$~cm$^{-3}$. 

The {\sc fruity} and Monash models are based on different stellar evolution codes, which assume different physical and nuclear inputs. Additionally, these models adopt different approaches to perform their detailed nucleosynthesis computations, mainly regarding to the formation of the $^{13}$C pocket, since the $^{13}$C left in the He-intershell is not able to provide the required neutron densities to reproduce the observations \citep[][]{busso1995, busso2001, abia2001}{}{}. To address this challenge, modelers assume the occurrence of a proton ingestion in the He-intershell at the time of TDU, which leads the formation of a $^{13}$C pocket in the next interpulse period and provides the neutron reservoir to the $s$-process \citep[][]{straniero1995, straniero1997}{}{}. However, this standard approach introduces the major uncertainty source in the predictions \citep[see, e.g.,][]{karakas2014}{}{}. The {\sc fruity} models self-consistently produce the $^{13}$C pocket from a time-dependent convective overshoot implementation \citep[][]{cristallo2009}. In the Monash models, a parametric approach leads to the $^{13}$C pocket formation, by artificially inserting a mass of protons (controlled by the $M_{\textrm{mix}}$ parameter) in the top layers of the He-intershell during the TDU \citep[][]{lugaro2012}. A detailed comparison between the {\sc fruity} and Monash models is presented by \citet[][]{karakas2016}{}{}.

\begin{figure*}
    \centering
    \includegraphics{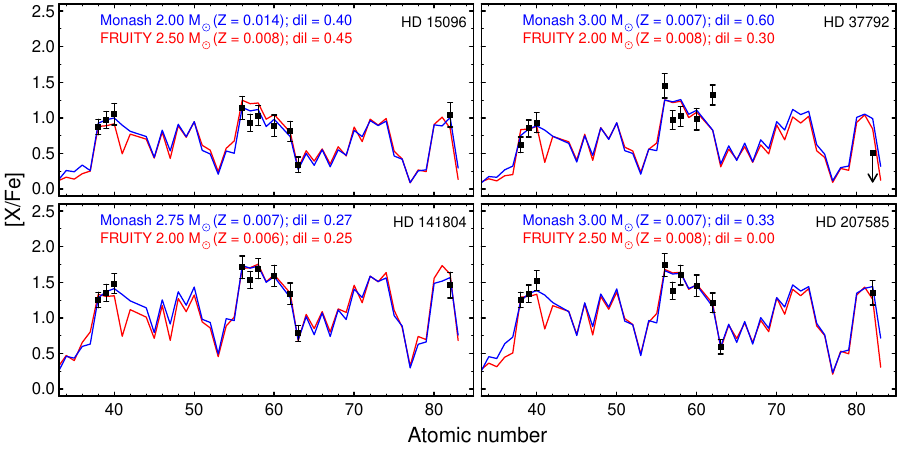}
    \caption{Comparison between the observed abundance profiles (black squares with error bars) and the predicted profiles (curves) from the Monash (blue) and {\sc fruity} (red) nucleosynthesis models that best fit the observational data set, after dilution is applied. The derived dilution factors are also shown.}
    \label{fig:comparison}
\end{figure*}

In this section, we perform a target-to-target comparison between the observed abundance patterns in our stars and those predicted by the (non rotating) {\sc fruity} and Monash models. For this, we have collected the [X/Fe] ratios predicted by these models, computed after the last TP, at the AGB surface. However, the $s$-processed material transferred was further diluted in the atmosphere of the observed Ba star. To take these effects into account, we must introduce a dilution factor ($dil$) in the predictions. This parameter is defined as the ratio $\log(M_{\textrm{Ba}}^{\textrm{env}}/M_{\textrm{AGB}}^{\textrm{transf}})$, where $M_{\textrm{Ba}}^{\textrm{env}}$ is the mass of the envelope of the Ba star after mass transfer and $M_{\textrm{AGB}}^{\textrm{transf}}$ is the mass transferred from the former AGB star. In this way, the diluted $\textrm{[X/Fe]}_{\textrm{Ba}}$ ratio expected to be observed in Ba stars is given by the parametric equation:

\begin{equation}
    \textrm{[X/Fe]}_{\textrm{Ba}}=\log\left[f \times 10^{\textrm{[X/Fe]}_{\textrm{ini}}}+10^{\textrm{[X/Fe]}_{\textrm{AGB}}-dil}\right],
\end{equation}

\noindent where $f=1-10^{dil}$, $\textrm{[X/Fe]}_{\textrm{ini}}$ is the initial (solar) composition adopted in the models and $\textrm{[X/Fe]}_{\textrm{AGB}}$ is the final abundance predicted by the models. The dilution factor has the effect of lowering the predicted abundances but without changing the shape of the distribution. Previous studies adopted that methodology in order to derive dilution factors and infer in the mass of the AGB progenitor star \citep[see, e.g.,][]{husti2009, shejeelammal2020, cseh2022}. We have applied different dilution factors to each model of our grid and compared the diluted predictions to the observed abundance patterns, evaluating the quality of the fits from the $\chi^{2}$ values. From a $\chi^{2}$ minimization, we find the models that best fit the observed data, and consequently the dilution factors.

In Figure~\ref{fig:comparison}, we show the neutron-capture patterns observed in HD~15096, HD~37792, HD~141804, and HD~207585, and the Monash and {\sc fruity} models that best fit the observations. We can see that $s$-process nucleosynthesis models for low-mass ($\leq 3.0$ M$_{\odot}$) AGB stars are able to reproduce successfully our observations. In their analysis, using the {\sc fruity} models, \citet{shejeelammal2020} fitted the abundance pattern of HD~207585 with a 2.5~M$_{\odot}$ ($Z=0.008$) model. Our results are consistent with many other studies that have evidenced the low-mass nature of the former polluter TP-AGB stars that contaminated the envelope of Ba stars \citep[e.g.,][]{allen2006, karinkuzhi2018b, cseh2018, cseh2022, shejeelammal2020, denhartog2023}{}{}. In connection, this favors the $^{13}$C as main neutron-source powering the $s$-process in low-mass AGB stars. Furthermore, if TP-AGB stars of higher masses were responsible for the patterns observed in the program stars, it would be expected to find magnesium excesses in their atmospheres, in consequence of the operation of the $^{22}$Ne neutron source. This is not observed in our stars, which exhibit normal [Mg/Fe] ratios (see Tables~\ref{tab:abd_hd37792_hd15096} and \ref{tab:abd_hd141804_hd207585}). Concerning the dilution factors, low values were found ($dil \leq 0.45$), which is reasonable, since Ba dwarfs do not present extended atmospheres. \citet{husti2009} reported $dil \lesssim 1.0$ for the Ba dwarfs of \citet{allen2006}. From the mass estimation of the polluted TP-AGB stars, we have applied the empirical initial–final mass relation of \citet{elbadry2018} to constraint the masses of WD companions of our stars. We found $\langle M_{\textrm{WD}}\rangle\sim0.66$ M$_{\odot}$, in agreement with values derived by \citet{escorza2023}, who combined astrometric data in their analysis.

\section{kinematics}\label{sec:kinematic}

The kinematics properties of the Ba dwarfs analyzed in this work were obtained following the methodology outlined in \citet{decastro2016}. Distances and proper motions were obtained based on kinematic Gaia EDR3 data \citep[][]{gaiaedr3}{}{} while radial velocities were obtained based on Doppler shift of the spectral absorption lines (Section~\ref{sec:radial_velocity}). Then we determined the spatial velocities relative to the local standard of rest, $U_{\rm LSR}$, $V_{\rm LSR}$, $W_{\rm LSR}$, where $U_{\rm LSR}$ is positive toward the Galactic center, $V_{\rm LSR}$ is positive in direction of Galactic rotation ($l$\,=\,90$^\circ$, $b$\,=\,0$^\circ$), and $W_{\rm LSR}$ is positive toward the north Galactic pole ($b$\,=\,90$^\circ$). We assumed the solar motion of (11.1, 12.2, 7.3) km\,s$^{-1}$, as derived by \citet{schonrich2010} and the algorithm of \citet{johnson1987}. Finally, the probability of a Ba dwarf star belonging to the thin disk, thick disc, or halo population was calculated, following the procedure described in \citet{ramirez2013}. Membership to a given population was established when the star had a probability $P_{\rm population}$ greater than or equal to 70\%. Table~\ref{tab:kinematic} shows the results obtained for the spatial velocities and the corresponding probabilities.

\begin{table*}
\centering
\caption{Kinematic data for the Ba dwarfs analyzed in this work. The radial velocities are given in the second column. The spacial velocities with their respective uncertainties are given from the third to fifth columns. The sixth, seventh, and eighth columns give the probability of a star to be a member of the thin disk ($P_{\textrm{thin}}$), thick disk ($P_{\textrm{thick}}$) and the halo ($P_{\textrm{halo}}$). The last column gives the spatial velocity of the star.}\label{tab:kinematic}
\begin{tabular}{lccccccc}\hline
Star        &  $RV$             & $U_{\rm LSR}$   & $V_{\rm LSR}$   & $W_{\rm LSR}$   & $P_{\textrm{thin}}$ & $P_{\textrm{thick}}$ & $P_{\textrm{halo}}$ \\
            &  km\,s$ ^{-1}$    & km\,s$ ^{-1}$   & km\,s$ ^{-1}$   & km\,s$ ^{-1}$   &      &      &      \\\hline
HD~15096    & $-$5.62$\pm$0.19  & $-$30.5$\pm$0.5 & $-$7.0$\pm$0.2  & $+$42.5$\pm$0.4 & 0.95 & 0.05 & 0.00 \\
HD~37792    & $+$10.13$\pm$0.44 & $+$56.0$\pm$1.5 & $-$10.7$\pm$0.6 & $-$56.5$\pm$1.7 & 0.76 & 0.23 & 0.00 \\
HD~141806   & $-$55.30$\pm$0.41 & $-$47.4$\pm$0.4 & $+$23.4$\pm$0.4 & $-$10.8$\pm$0.3 & 0.98 & 0.02 & 0.00 \\
HD~207585   & $-$55.89$\pm$0.31 & $-$21.6$\pm$0.2 & $-$39.7$\pm$1.4 & $+$35.8$\pm$0.6 & 0.88 & 0.12 & 0.00 \\\hline
\end{tabular}
\end{table*}

\begin{figure}
    \centering
    \includegraphics[width=\columnwidth]{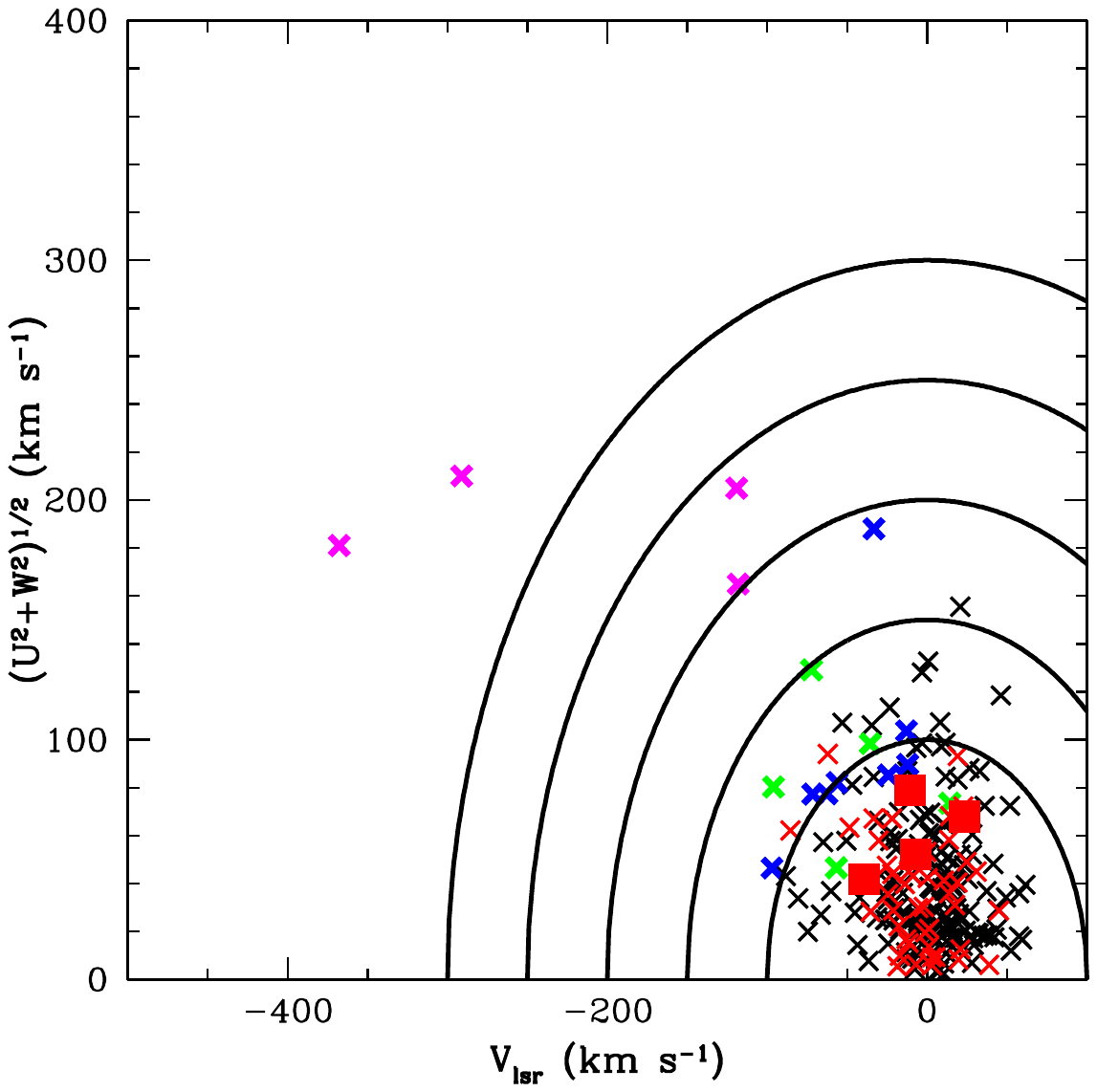}
    \caption{The Ba dwarfs analyzed in this work (red squares) in the Toomre diagram of $(U^{2}_{\rm LSR}+W^{2}_{\rm LSR})^{1/2}$ {\sl versus} $V_{\rm LSR}$. Ba dwarfs previously analyzed by \citet{escorza2019}, \citet{pereira2011a}, and \citet{pereira2005} are shown in red crosses. Classical Ba giant stars of the thin disk (black crosses), transition thin/thick disk (green crosses), thick disk (blue crosses), and halo (magenta crosses), taken from \citet{decastro2016}, are also plotted in this diagram.}
    \label{fig:toomre}
\end{figure}

Figure~\ref{fig:toomre} display the Toomre diagram of $(U^{2}_{\rm LSR}+W^{2}_{\rm LSR})^{1/2}$ {\sl versus} $V_{\rm LSR}$, where the stars are kinematically classified according to their spatial velocities and probabilities. In addition to the program stars, the position of some Ba dwarf stars analyzed by \citet{escorza2019}, \citet{pereira2011a}, and \citet{pereira2005} are also shown. Data for Ba giants are included in Figure~\ref{fig:toomre}. HD~15096, HD~37792, HD~141804, and HD~207585 exhibit kinematical properties consistent with thin disk stars, as indicated by their membership probabilities. Most of the previously analyzed Ba dwarfs also belong to the thin disk, with one exception for the star HD~6434, which can be considered as a thin-thick disc star with $P_{\rm thin}$ and $P_{\rm thick}$ probabilities of 33\% and 66\%, respectively.

\section{Conclusions}\label{sec:summary}

We conducted a classical LTE analysis based on high-resolution spectroscopic data for a sample of four chemically peculiar stars. HD~15096, HD~37792, and HD~141804 were considered as potential Ba dwarf candidates, whereas HD~207585 is a known Ba dwarf star. We have determined their atmospheric parameters, from clean and sufficiently unblended Fe\,{\sc i} and Fe\,{\sc ii} absorption lines, and compared them with values previously reported in the literature. We found effective temperatures between $5300-6500$ K, surface gravities within the interval $3.90\leq \log g\leq 4.50$, and $-0.55\leq\rm{[Fe/H]}\leq +0.14$. 

Subsequently, chemical abundances were extracted for a set of 21 elements, including the CNO group and neutron-capture elements. We found moderate carbon excesses in these stars, with [C/Fe] ratios ranging from $+0.29$ to $+0.66$, and that carbon is mainly responsible for the CNO excesses observed in their atmospheres. HD~15096 and HD~37792 show $\rm{C/O}<1$, typical for Ba stars, while HD~141804 and HD~207585 show $\rm{C/O}>1$, typical for CH stars. For the elements from sodium to nickel, these systems follow the Galactic trend. On the other hand, as far as the $s$-process elements are concerned, the program stars show high levels of enrichment, with [$s$/Fe]~$\gtrsim +1.00$ dex. All these stars are identified as spectroscopic binaries in the SIMBAD database, and orbital elements are provided in the literature for HD~15096, HD~141804, and HD~207585. Hence, their chemical peculiarities are attributable to mass transfer events. We applied the classical approach of the $s$-process (the $\sigma N_{s}$ curve) to estimate the neutron exposure of the $s$-processed material observed in their envelopes, yielding $\tau_{0}$ of the order of 0.6 to 0.7 mb$^{-1}$. 

We individually compared the observed abundance patterns in our stars with Monash and {\sc fruity} nucleosynthesis models of the $s$-process. Consequently, we were able to estimate dilution factors and masses of the former polluting TP-AGB stars. Low-mass ($M\lesssim 3.0~\rm{M}_{\odot}$) models successfully reproduced our observations. Notably, no Na enrichment was detected in the program stars, providing additional evidence for the operation of the $^{13}$C as main neutron source of the TP-AGB stars. We applied the empirical initial–final mass relation of \citet{elbadry2018} to estimate the masses of the WD companions of our stars, yielding $\langle M_{\textrm{WD}}\rangle\sim0.66$ M$_{\odot}$, which is consistent for Ba and related stars. From the kinematic point of view, we estimated the probabilities of these stars belonging to thin disc, thick disc, and halo. We found that these objects are members of thin disc, with probabilities greater than 70\%. 

In conclusion, we have identified HD~15096 and HD~37792 as two new Ba dwarfs, and confirmed the Ba dwarf nature of HD~141804. As we pointed out, the current sample of Ba dwarfs confirmed from high-resolution spectroscopic data is much smaller than the sample of the classical giants. Finally, we stressed on the need of identifying new Ba dwarf star candidates and exploring their chemical patterns.

\section*{Acknowledgments}
This work has been developed under a fellowship of the PCI Program of the Ministry of Science, Technology and Innovation - MCTI, financed by the Brazilian National Council of Research - CNPq, through the grant 300438/2024-9. NH acknowledges the fellowships (300181/2023-0 and 300434/2024-3) of the PCI Program - MCTI and CNPq. NAD acknowledges Funda\c{c}\~ao de Amparo \`a Pesquisa do Estado do Rio de Janeiro - FAPERJ, Rio de Janeiro, Brazil, for grant E-26/203.847/2022. AS acknowledges CNPq for a PhD fellowship 141219/2023-8. This work has made use of the {\sc vald} database, operated at Uppsala University, the Institute of Astronomy RAS in Moscow, and the University of Vienna. This research has made use of NASA’s Astrophysics Data System Bibliographic Services.

\vspace{5mm}
\facilities{}
\software{{\sc iraf} \citep{tody1986}; {\sc moog} \citep{sneden1973, sneden2012}; R and RStudio \citep{rrstudio}}

\bibliographystyle{aasjournal}
\bibliography{references.bib}



\appendix

\setcounter{table}{0}

\renewcommand{\thetable}{A\arabic{table}}
\renewcommand*{\theHtable}{\thetable}
\begin{longtable}{lccccccc}
\caption{Equivalent width measurements of Fe\,{\sc i} and Fe\,{\sc ii} lines.}\label{app:ew_iron}\\
    \toprule
                      &            &              &                    & \multicolumn{4}{c}{Equivalent widths (m\AA)} \\
                                                                         \cline{5-8}
        Element       & Wavelength & $\chi$\,(eV) & $\log$ \textit{gf} & HD\,15096 & HD\,37792 & HD\,141804 & HD\,207585\\
        
    \endfirsthead
    
    \multicolumn{8}{c}%
    {{\tablename\ \thetable{} -- Continued}} \\
    
    \toprule
                      &            &              &                    & \multicolumn{4}{c}{Equivalent widths (m\AA)} \\
                                                                         \cline{5-8}
        Element       & Wavelength & $\chi$\,(eV) & $\log$ \textit{gf} & HD\,15096 & HD\,37792 & HD\,141804 & HD\,207585\\
        
    \midrule
    
    \endhead
    
    \midrule
    
    \multicolumn{8}{r}{{\textit{Continued on next page}}} \\
    
    \endfoot
    
    \bottomrule

    \endlastfoot
    
    \midrule

Fe\,{\sc i} &  5\,133.69  & 4.18  & $+$0.20   & --- & --- & 108 & --- \\
            &  5\,150.84  & 0.99  & $-$3.00   & 135 & 58  &  75 & --- \\
            &  5\,151.91  & 1.01  & $-$3.32   & --- & 41  & --- & --- \\
            &  5\,159.06  & 4.28  & $-$0.65   &  77 & 28  &  39 &  53 \\
            &  5\,162.27  & 4.18  & $+$0.07   & --- & 80  & --- & --- \\
            &  5\,194.94  & 1.56  & $-$2.09   & --- & 80  &  91 & 109 \\
            &  5\,198.71  & 2.22  & $-$2.14   & 111 & 49  &  63 &  76 \\
            &  5\,232.94  & 2.94  & $-$0.08   & --- & 117 & --- & --- \\
            &  5\,242.49  & 3.63  & $-$0.97   &  93 & 48  &  54 &  71 \\
            &  5\,250.21  & 0.12  & $-$4.92   &  80 & --- & --- &  50 \\
            &  5\,281.79  & 3.04  & $-$0.83   & --- & 83  &  91 & 108 \\
            &  5\,288.52  & 3.69  & $-$1.51   &  65 & 18  &  27 &  40 \\
            &  5\,302.31  & 3.28  & $-$0.74   & --- & 72  & --- &  ---\\
            &  5\,307.36  & 1.61  & $-$2.97   &  96 & 38  &  54 &   70\\
            &  5\,321.11  & 4.43  & $-$1.19   &  49 & --- & --- &  ---\\
            &  5\,322.04  & 2.28  & $-$2.84   &  68 & --- &  26 &   40\\
            &  5\,339.93  & 3.27  & $-$0.68   & --- & 76  &  90 &  102\\
            &  5\,341.02  & 1.61  & $-$1.95   & --- & 87  & 103 &  128\\
            &  5\,353.37  & 4.10  & $-$0.68   &  95 & --- & --- &  ---\\
            &  5\,364.87  & 4.45  & $+$0.23   & --- & 73  &  80 &   93\\
            &  5\,367.47  & 4.42  & $+$0.43   & --- & 76  & --- &  103\\
            &  5\,373.71  & 4.47  & $-$0.71   &  66 & 25  &  32 &   46\\
            &  5\,389.48  & 4.42  & $-$0.25   & --- & 41  &  54 &   64\\
            &  5\,393.17  & 3.24  & $-$0.72   & --- & 72  & --- &  ---\\
            &  5\,400.50  & 4.37  & $-$0.10   & 125 & 59  &  79 & --- \\
            &  5\,405.77  & 0.99  & $-$1.85   & --- & 111 & --- & --- \\
            &  5\,410.91  & 4.47  & $+$0.40   & --- & 73  & --- & 109 \\
            &  5\,417.03  & 4.42  & $-$1.53   &  35 & --- & --- & --- \\
            &  5\,434.52  & 1.01  & $-$2.12   & --- & 103 & --- & --- \\
            &  5\,441.34  & 4.31  & $-$1.58   &  38 & --- & --- & --- \\
            &  5\,445.04  & 4.39  & $+$0.04   & --- & 64  & --- &  86 \\
            &  5\,446.92  & 0.99  & $-$1.91   & --- & 112 & --- & --- \\
            &  5\,487.75  & 4.32  & $-$0.65   &  86 & 41  & --- & --- \\
            &  5\,506.78  & 0.99  & $-$2.80   & --- & 70  & --- &  105\\
            &  5\,522.45  & 4.21  & $-$1.40   &  47 & --- & --- &   32\\
            &  5\,532.75  & 3.57  & $-$2.00   &  50 & 14  & --- &  ---\\
            &  5\,554.90  & 4.55  & $-$0.38   &  92 & 43  &  52 &  ---\\
            &  5\,560.21  & 4.43  & $-$1.04   &  52 & 16  &  22 &   32\\
            &  5\,563.60  & 4.19  & $-$.840   &  81 & 31  &  49 &  ---\\
            &  5\,567.39  & 2.61  & $-$2.56   &  73 & --- &  29 &  ---\\
            &  5\,569.62  & 3.42  & $-$0.49   & --- & 74  &  94 &  ---\\
            &  5\,572.84  & 3.40  & $-$0.28   & --- & 89  & --- &  ---\\
            &  5\,576.09  & 3.43  & $-$0.85   & --- & 58  &  74 &  ---\\
            &  5\,584.77  & 3.57  & $-$2.17   &  47 & --- & --- &  ---\\
            &  5\,624.02  & 4.39  & $-$1.33   &  39 & --- & --- & --- \\
            &  5\,633.95  & 4.99  & $-$0.12   &  78 & 29  & --- & --- \\
            &  5\,635.82  & 4.26  & $-$1.74   &  32 & --- & --- & --- \\
            &  5\,638.26  & 4.22  & $-$0.72   &  79 & --- &  45 &  60 \\
            &  5\,686.53  & 4.55  & $-$0.45   &  79 & --- & --- &  56 \\
            &  5\,691.50  & 4.30  & $-$1.37   &  43 & --- &  22 &  30 \\
            &  5\,705.47  & 4.30  & $-$1.36   &  42 & --- & --- &  22 \\
            &  5\,717.83  & 4.28  & $-$0.97   &  65 & --- &  30 &  44 \\
            &  5\,731.76  & 4.26  & $-$1.15   &  58 & 16  &  28 &  43 \\
            &  5\,762.99  & 4.21  & $-$0.41   & 109 & 61  & --- &  78 \\
            &  5\,806.73  & 4.61  & $-$0.90   &  51 & 15  &  22 &  33 \\
            &  5\,814.81  & 4.28  & $-$1.82   &  22 & --- & --- &  14 \\
            &  5\,852.22  & 4.55  & $-$1.18   &  42 & --- &  16 &  23 \\
            &  5\,883.82  & 3.96  & $-$1.21   &  69 & 19  &  32 &  47 \\
            &  5\,916.25  & 2.45  & $-$2.99   &  60 & 10  & --- &  37 \\
            &  5\,934.65  & 3.93  & $-$1.02   &  83 & 25  &  41 &  64 \\
            &  6\,016.66  & 3.55  & $-$1.67   &  73 & --- & --- & --- \\
            &  6\,024.06  & 4.55  & $-$0.06   & --- & 57  &  71 &  80 \\
            &  6\,027.05  & 4.08  & $-$1.09   &  67 & 22  &  37 &  49 \\
            &  6\,056.01  & 4.73  & $-$0.40   &  73 & 29  &  41 &  48 \\
            &  6\,065.48  & 2.61  & $-$1.53   & 129 & 63  &  77 &  92 \\
            &  6\,079.01  & 4.65  & $-$0.97   &  44 & 14  & --- &  27 \\
            &  6\,082.71  & 2.22  & $-$3.58   &  39 & --- & --- & --- \\
            &  6\,093.64  & 4.61  & $-$1.35   &  31 & --- & --- &  17 \\
            &  6\,096.66  & 3.98  & $-$1.78   &  38 & --- & --- &  18 \\
            &  6\,136.61  & 2.45  & $-$1.40   & --- & 80  &  90 & 103 \\
            &  6\,137.69  & 2.59  & $-$1.40   & --- & 71  &  87 & 103 \\
            &  6\,151.62  & 2.18  & $-$3.29   &  57 & --- &  18 &  30 \\
            &  6\,157.73  & 4.08  & $-$1.11   &  58 & 23  &  34 & --- \\
            &  6\,165.36  & 4.14  & $-$1.47   &  45 & 16  &  18 &  27 \\
            &  6\,170.51  & 4.79  & $-$0.38   &  78 & 31  &  42 &  55 \\
            &  6\,173.34  & 2.22  & $-$2.88   &  74 & 20  &  30 &  49 \\
            &  6\,187.99  & 3.94  & $-$1.57   &  51 & --- &  20 &  28 \\
            &  6\,191.56  & 2.43  & $-$1.42   & --- & 75  &  87 & 107 \\ 
            &  6\,200.31  & 2.60  & $-$2.44   &  75 & 20  &  34 &  52 \\
            &  6\,213.43  & 2.22  & $-$2.48   &  91 & 30  &  47 &  62 \\
            &  6\,230.72  & 2.56  & $-$1.28   & --- & 83  &  96 & 109 \\
            &  6\,252.56  & 2.40  & $-$1.72   & 130 & 66  &  82 &  97 \\
            &  6\,254.26  & 2.28  & $-$2.44   &  97 & --- & --- & --- \\
            &  6\,265.13  & 2.18  & $-$2.55   &  96 & 33  &  50 &  70 \\
            &  6\,311.50  & 2.83  & $-$3.23   &  32 & --- & --- & --- \\
            &  6\,322.69  & 2.59  & $-$2.43   &  78 & 23  &  35 & --- \\
            &  6\,380.74  & 4.19  & $-$1.32   &  53 & 15  &  25 &  33 \\
            &  6\,392.54  & 2.28  & $-$4.03   &  21 & --- & --- & --- \\
            &  6\,393.60  & 2.43  & $-$1.43   & --- & 71  &  88 & 102 \\
            &  6\,411.65  & 3.65  & $-$0.66   & --- & 61  &  77 &  93 \\
            &  6\,419.95  & 4.73  & $-$0.09   &  89 & 36  &  49 &  59 \\
            &  6\,421.35  & 2.28  & $-$2.01   & 122 & 61  &  77 &  94 \\
            &  6\,430.85  & 2.18  & $-$2.01   & 124 & 61  &  74 &  93 \\
            &  6\,436.41  & 4.19  & $-$2.46   &  12 & --- & --- & --- \\
            &  6\,469.19  & 4.83  & $-$0.62   &  56 & 16  &  29 &  40 \\
            &  6\,518.37  & 2.83  & $-$2.30   &  71 & 14  & --- & --- \\
            &  6\,551.68  & 0.99  & $-$5.79   &  11 & --- & --- & --- \\
            &  6\,574.23  & 0.99  & $-$5.02   &  39 & --- & --- &  17 \\
            &  6\,592.91  & 2.72  & $-$1.47   & 126 & 54  & --- &  94 \\
            &  6\,593.87  & 2.44  & $-$2.42   &  89 & 33  &  46 &  67 \\
            &  6\,597.56  & 4.79  & $-$0.92   &  45 & --- &  17 &  29 \\
            &  6\,608.03  & 2.28  & $-$4.03   &  23 & --- & --- & --- \\
            &  6\,609.11  & 2.56  & $-$2.69   &  69 & 15  &  29 &  46 \\
            &  6\,646.93  & 2.61  & $-$3.99   &  14 & --- & --- & --- \\
            &  6\,699.14  & 4.59  & $-$2.19   &  10 & --- & --- & --- \\    
            &  6\,703.57  & 2.76  & $-$3.16   &  38 & --- & --- &  22 \\
            &  6\,710.32  & 1.80  & $-$4.88   &  14 & --- & --- & --- \\
            &  6\,713.74  & 4.79  & $-$1.60   &  19 & --- & --- &  11 \\
            &  6\,739.52  & 1.56  & $-$4.95   &  58 & --- & --- &  30 \\
            &  6\,750.15  & 2.42  & $-$2.62   &  82 & 27  &  38 &  56 \\
            &  6\,752.71  & 4.64  & $-$1.20   &  39 & --- &  18 &  26 \\
            &  6\,783.70  & 2.59  & $-$3.98   &  15 & --- & --- & --- \\
            &  6\,793.26  & 4.07  & $-$2.47   &  14 & --- & --- & --- \\
            &  6\,806.85  & 2.73  & $-$3.21   &  40 & --- & --- &  17 \\    
            &  6\,810.26  & 4.61  & $-$0.99   &  49 & 14  & --- &  26 \\
            &  6\,820.37  & 4.64  & $-$1.17   &  41 & --- &  17 &  25 \\
            &  6\,841.34  & 4.61  & $-$0.60   &  70 & 22  &  32 & --- \\
            &  6\,858.15  & 4.61  & $-$0.93   &  51 & --- &  28 & --- \\\midrule
 
 Fe\,{\sc ii}   &  4\,993.35   & 2.81  & $-$3.67  &  31 & 25  &  28 & 41  \\
                &  5\,132.66   & 2.81  & $-$4.00  &  17 & --- & --- & --- \\
                &  5\,197.56   & 3.23  & $-$2.25  &  64 & --- & --- & --- \\
                &  5\,234.62   & 3.22  & $-$2.24  &  66 & 84  &  74 &  83 \\
                &  5\,284.10   & 2.89  & $-$3.01  &  51 & --- &  49 &  63 \\
                &  5\,325.56   & 3.22  & $-$3.17  &  29 & 34  &  32 &  38 \\
                &  5\,414.05   & 3.22  & $-$3.62  &  16 & 19  &  20 & --- \\
                &  5\,425.25   & 3.20  & $-$3.21  &  31 & 33  &  34 &  49 \\
                &  5\,534.83   & 3.25  & $-$2.77  &  43 & 55  &  45 & --- \\
                &  5\,991.37   & 3.15  & $-$3.56  &  21 & 21  &  22 &  30 \\
                &  6\,084.10   & 3.20  & $-$3.80  &  12 & 15  &  14 &  21 \\
                &  6\,149.25   & 3.89  & $-$2.72  &  22 & 29  &  26 &  35 \\   
                &  6\,247.55   & 3.89  & $-$2.34  &  34 & 52  &  47 &  53 \\
                &  6\,416.92   & 3.89  & $-$2.68  &  27 & 27  &  32 &  33 \\
                &  6\,432.68   & 2.89  & $-$3.58  & --- & 32  & --- &  38 \\
\end{longtable}

\renewcommand\thetable{A\arabic{table}}
\renewcommand*{\theHtable}{\thetable}
\begin{longtable}{lcccccccc}
\caption{Atomic line lists for other chemical species.}\label{app:ew_other}\\

    \toprule    
                 &             &              &                    &      &  \multicolumn{4}{c}{Equivalent widths (m\AA)} \\
                                                                             \cline{6-9}
        Element  &  Wavelength & $\chi$\,(eV) & $\log$ \textit{gf} &  Ref & HD\,15096 & HD\,37792 & HD\,141804 & HD\,207585 \\
        
    \endfirsthead

    \multicolumn{9}{c}%
    {{\tablename\ \thetable{} -- Continued}} \\

    \toprule
    
                 &             &              &                    &      &  \multicolumn{4}{c}{Equivalent widths (m\AA)} \\
                                                                             \cline{6-9}
        Element  &  Wavelength & $\chi$\,(eV) & $\log$ \textit{gf} &  Ref & HD\,15096 & HD\,37792 & HD\,141804 & HD\,207585 \\
        
    \midrule
        
    \endhead

    \midrule
    
    \multicolumn{9}{r}{{\textit{Continued on next page}}} \\
    
    \endfoot

    \bottomrule
    
    \endlastfoot

    \midrule

C\,{\sc i}  & 4\,770.03  &  7.48  & $-$2.33  & T02     & ---  & 17  &  22 &  21 \\
            & 4\,775.90  &  7.49  & $-$2.19  & T02     &  11  & 22  &  30 & --- \\
            & 4\,932.05  &  7.69  & $-$1.66  & T02     & ---  & 33  &  53 & --- \\
            & 5\,052.10  &  7.68  & $-$1.30  & T02     &  26  & 49  &  67 &  70 \\
            & 5\,380.34  &  7.68  & $-$1.84  & T02     & ---  & 33  &  43 &  45 \\
            & 6\,587.60  &  8.54  & $-$1.22  & L82     & ---  & 27  & --- &  38 \\
            & 7\,111.48  &  8.64  & $-$1.32  & L82     & ---  & 17  & --- &  24 \\
            & 7\,113.18  &  8.64  & $-$0.95  & L82     & ---  & 30  &  43 &  37 \\
            & 7\,115.19  &  8.64  & $-$0.90  & L82     & ---  & 30  &  40 &  47 \\
            & 7\,116.99  &  8.64  & $-$1.08  & L82     & ---  & 25  &  32 &  37 \\
            & 9\,061.48  &  7.48  & $-$0.34  & TS97    & ---  & --- & 177 & --- \\
            & 9\,078.32  &  7.48  & $-$0.58  & TS97    & ---  & 160 & --- & 156 \\ 
            & 9\,111.85  &  7.49  & $-$0.29  & TS97    & 105  & --- & --- & --- \\\midrule

O\,{\sc i}  & 7\,771.94  &  9.15  & $+$0.32  & TH05    & ---  & 126 &  89 &  73 \\
            & 7\,774.16  &  9.15  & $+$0.17  & TH05    & ---  & 111 &  75 &  68 \\
            & 7\,775.35  &  9.15  & $-$0.04  & TH05    & ---  & 88  &  63 &  61 \\\midrule

Na\,{\sc i} & 5\,682.65  &  2.10  & $-$0.70   & PS     & 96   & 48  &  63 &  79 \\
            & 5\,688.22  &  2.10  & $-$0.40   & PS     & 125  & 68  &  86 & 104 \\
            & 6\,154.22  &  2.10  & $-$1.57   & R03    & 39   & 13  &  15 &  24 \\
            & 6\,160.75  &  2.10  & $-$1.27   & R03    & 65   & 19  &  25 &  39 \\\midrule 

Mg\,{\sc i} & 4\,571.10  &  0.00  & $-$5.57   & N96    & 124  & 46  &  67 &  86 \\ 
            & 4\,702.99  &  4.35  & $-$0.38   & N96    & ---  & 144 & --- & --- \\
            & 4\,730.04  &  4.34  & $-$2.39   & R03    &  78  & 25  &  35 &  41 \\
            & 5\,528.42  &  4.34  & $-$0.36   & J2006  & ---  & 139 & --- & --- \\
            & 5\,711.10  &  4.34  & $-$1.68   & R99    & 118  & 54  &  66 &  81 \\
            & 6\,318.71  &  5.11  & $-$1.94   & Ca2007 & ---  & 10  & --- & --- \\
            & 8\,712.69  &  5.93  & $-$1.26   & Ca2007 &  65  & --- & --- & --- \\
            & 8\,717.83  &  5.91  & $-$0.97   & WSM    &  84  & --- & --- & --- \\
            & 8\,736.04  &  5.94  & $-$0.34   & WSM    & ---  & 78  &  91 & --- \\\midrule

Al\,{\sc i} & 6\,696.03  &  3.14  & $-$1.48   & MR94   &  39  & --- &  11 &  19 \\
            & 6\,698.67  &  3.14  & $-$1.63   & R03    &  29  & --- & --- &  11 \\
            & 7\,835.32  &  4.04  & $-$0.580  & R03    &  42  & --- &  19 &  20 \\
            & 7\,836.13  &  4.02  & $-$0.400  & R03    &  57  & --- &  21 &  25 \\
            & 8\,772.88  &  4.02  & $-$0.250  & R03    &  79  & --- & --- & --- \\
            & 8\,773.91  &  4.02  & $-$0.070  & R03    &  93  & --- & --- & --- \\\midrule

Si\,{\sc i} & 5\,793.08  &  4.93  & $-$2.06   & R03    &  37  & --- & --- &  32 \\
            & 6\,125.03  &  5.61  & $-$1.54   & E93    &  29  & 12  &  19 &  25 \\
            & 6\,131.58  &  5.62  & $-$1.69   & E93    &  21  & --- &  13 & --- \\
            & 6\,145.08  &  5.62  & $-$1.43   & E93    &  33  & 12  &  20 &  26 \\
            & 6\,155.14  &  5.62  & $-$0.77   & E93    & ---  & 40  & --- & --- \\
            & 8\,728.01  &  6.18  & $-$0.36   & E93    & ---  & 47  & --- & --- \\
            & 8\,742.45  &  5.87  & $-$0.51   & E93    &  89  & 60  & --- & --- \\\midrule

Ca\,{\sc i} & 5\,581.80  &  2.52  & $-$0.67   & C2003  & 104  & 47  &  63 &  77 \\
            & 5\,601.29  &  2.52  & $-$0.52   & C2003  & 125  & --- &  70 &  88 \\
            & 5\,857.46  &  2.93  & $+$0.11   & C2003  & 145  & 79  &  87 & 100 \\
            & 5\,867.57  &  2.93  & $-$1.61   & C2003  & ---  & --- & --- &  12 \\
            & 6\,102.73  &  1.88  & $-$0.79   & D2002  & ---  & --- &  86 &  99 \\
            & 6\,122.23  &  1.89  & $-$0.32   & D2002  & ---  & 101 & 121 & 132 \\
            & 6\,161.30  &  2.52  & $-$1.27   & E93    & ---  & 18  &  26 &  43 \\
            & 6\,162.18  &  1.90  & $-$0.09   & D2002  & ---  & 119 & 130 & 151 \\
            & 6\,166.44  &  2.52  & $-$1.14   & R03    &  77  & 29  &  34 &  46 \\
            & 6\,169.04  &  2.52  & $-$0.80   & R03    & 103  & 40  &  53 &  64 \\
            & 6\,169.56  &  2.53  & $-$0.48   & DS91   & 123  & 55  &  71 &  81 \\
            & 6\,439.08  &  2.52  & $+$0.47   & D2002  & ---  & 106 & --- & --- \\
            & 6\,449.82  &  2.52  & $-$0.50   & C2003  & ---  & 56  &  79 &  78 \\
            & 6\,455.60  &  2.51  & $-$1.29   & R03    &  63  & 17  &  23 &  37 \\
            & 6\,464.68  &  2.52  & $-$2.42   & C2003  &  15  & --- & --- & --- \\
            & 6\,471.66  &  2.51  & $-$0.69   & S86    &  96  & 49  &  64 &  69 \\
            & 6\,493.79  &  2.52  & $-$0.11   & DS91   & ---  & 80  &  88 & 100 \\
            & 6\,499.65  &  2.52  & $-$0.81   & C2003  &  92  & 39  &  52 &  65 \\
            & 6\,717.69  &  2.71  & $-$0.52   & C2003  & ---  & 52  &  66 &  82 \\\midrule

Ti\,{\sc i} & 4\,512.740  & 0.840 & $-$0.400  & L13    &  80  & --- & 33  & 52  \\
            & 4\,518.030  & 0.830 & $-$0.250  & L13    &  87  & 25  & 38  & 54  \\
            & 4\,533.249  & 0.850 & $+$0.540  & L13    &  --- & 61  & 72  & 85  \\
            & 4\,534.785  & 0.840 & $+$0.350  & L13    &  --- & 50  & 67  & 80  \\
            & 4\,548.770  & 0.830 & $-$0.281  & L13    &  80  & 22  & 37  & 54  \\
            & 4\,555.490  & 0.850 & $-$0.400  & L13    &  74  & 20  & 36  & --- \\
            & 4\,617.280  & 1.750 & $+$0.439  & L13    &  --- & 18  & 32  & 46  \\
            & 4\,639.950  & 1.740 & $-$0.190  & L13    &  54  & --- & 13  & 26  \\
            & 4\,758.120  & 2.250 & $+$0.510  & L13    &  54  & 14  & 16  & 28  \\
            & 4\,759.280  & 2.250 & $+$0.590  & L13    &  55  & 12  & 19  & 29  \\
            & 4\,778.260  & 2.240 & $-$0.351  & L13    &  21  & --- & --- & --- \\
            & 4\,981.718  & 0.840 & $+$0.570  & L13    &  --- & 68  & --- & --- \\
            & 4\,981.740  & 0.850 & $+$0.570  & L13    &  --- & --- & 80  & 98  \\
            & 5\,016.170  & 0.850 & $-$0.480  & L13    &  --- & 14  & 29  & --- \\
            & 5\,022.870  & 0.830 & $-$0.331  & L13    &  --- & 25  & --- & 56  \\
            & 5\,039.960  & 0.020 & $-$1.080  & L13    &  92  & --- & --- & --- \\
            & 5\,043.590  & 0.840 & $-$1.590  & L13    &  33  & --- & --- & --- \\
            & 5\,145.470  & 1.460 & $-$0.539  & L13    &  56  & --- & --- & 24  \\
            & 5\,152.190  & 0.020 & $-$1.950  & L13    &  --- & --- & --- & 17  \\
            & 5\,173.750  & 0.000 & $-$1.060  & L13    &  --- & 28  & --- & --- \\
            & 5\,210.390  & 0.050 & $-$0.821  & L13    &  --- & 32  & 53  & 74  \\
            & 5\,219.710  & 0.020 & $-$2.220  & L13    &  43  & --- & --- & --- \\
            & 5\,295.780  & 1.050 & $-$1.590  & L13    &  23  & --- & --- & --- \\
            & 5\,490.160  & 1.460 & $-$0.840  & L13    &  33  & --- & --- & --- \\
            & 5\,866.460  & 1.070 & $-$0.790  & L13    &  65  & 10  & 17  & 26  \\
            & 5\,922.120  & 1.050 & $-$1.380  & L13    &  32  & --- & --- & 10  \\
            & 6\,091.180  & 2.270 & $-$0.320  & L13    &  20  & --- & --- & --- \\
            & 6\,126.220  & 1.050 & $-$1.370  & R03    &  36  & --- & --- & --- \\
            & 6\,258.110  & 1.440 & $-$0.390  & L13    &  61  & --- & 17  & 31  \\
            & 6\,261.100  & 1.430 & $-$0.530  & L13    &  63  & --- & 15  & 27  \\\midrule

Cr\,{\sc i} & 4\,254.346  & 0.000 & $-$0.090  & S07    & ---  & 127 & --- &  --- \\
            & 4\,289.729  & 0.000 & $-$0.370  & S07    & ---  & --- & 141 &  --- \\
            & 4\,789.340  & 2.540 & $-$0.330  & S07    & ---  & --- &  38 &  --- \\
            & 4\,936.340  & 3.110 & $-$0.250  & S07    & ---  &  12 &  17 &  --- \\
            & 5\,206.044  & 0.940 & $+$0.020  & S07    & ---  & --- & 123 &  --- \\
            & 5\,247.570  & 0.960 & $-$1.590  & S07    & ---  &  22 &  38 &   60 \\
            & 5\,296.690  & 0.982 & $-$1.360  & S07    & 104  &  34 & --- &   75 \\
            & 5\,300.750  & 0.980 & $-$2.000  & S07    & ---  &  13 & --- &  --- \\
            & 5\,345.800  & 1.003 & $-$0.950  & S07    & 127  &  56 &  77 &  --- \\
            & 5\,348.330  & 1.000 & $-$1.210  & S07    & 104  &  38 & --- &   73 \\
            & 5\,409.800  & 1.030 & $-$0.670  & S07    & ---  &  67 &  84 &  106 \\
            & 5\,787.930  & 3.322 & $-$0.080  & R03    &  50  & --- &  21 &   27 \\
            & 6\,330.097  & 0.940 & $-$2.920  & R03    &  39  & --- & --- &  --- \\\midrule

Ni\,{\sc i} & 4\,519.982  & 1.676  & $-$3.080  & W14   &  38  & --- & --- & --- \\
            & 4\,604.988  & 3.479  & $-$0.240  & W14   &  84  & --- & --- & --- \\
            & 4\,756.515  & 3.479  & $-$0.270  & W14   &  76  & --- & --- & --- \\
            & 4\,866.271  & 3.538  & $-$0.220  & W14   &  75  & --- & --- & --- \\
            & 4\,953.210  & 3.740  & $-$0.580  & W14   &  53  &  19 &  29 &  37 \\
            & 4\,976.326  & 1.676  & $-$3.000  & W14   &  43  & --- & --- & --- \\
            & 5\,010.940  & 3.635  & $-$0.979  & W14   & ---  &  17 &  22 &  33 \\
            & 5\,084.110  & 3.680  & $-$0.180  & E93   &  88  &  47 & --- &  70 \\
            & 5\,115.400  & 3.834  & $-$0.280  & R03   &  66  &  36 &  45 & --- \\
            & 5\,578.730  & 1.677  & $-$2.830  & W14   & ---  &  13 &  22 & --- \\
            & 5\,587.870  & 1.935  & $-$2.390  & W14   &  57  & --- & --- &  46 \\
            & 5\,748.360  & 1.677  & $-$3.240  & W14   &  31  & --- & --- & --- \\
            & 6\,108.120  & 1.677  & $-$2.600  & W14   &  62  & --- & --- &  46 \\
            & 6\,128.980  & 1.677  & $-$3.429  & W14   &  25  & --- & --- & --- \\
            & 6\,176.820  & 4.089  & $-$0.264  & R03   &  58  &  21 &  29 &  43 \\
            & 6\,177.250  & 1.826  & $-$3.460  & W14   &  15  & --- & --- & --- \\
            & 6\,204.610  & 4.089  & $-$1.080  & W14   &  19  & --- & --- & --- \\
            & 6\,223.990  & 4.106  & $-$0.910  & W14   &  22  & --- &  10 & --- \\
            & 6\,327.600  & 1.677  & $-$3.170  & W14   &  42  & --- &  13 &  24 \\
            & 6\,532.873  & 1.935  & $-$3.350  & W14   &  22  & --- & --- & --- \\
            & 6\,586.330  & 1.950  & $-$2.780  & W14   &  43  & --- &  14 &  24 \\
            & 6\,643.640  & 1.676  & $-$2.220  & W14   & ---  & --- &  46 & --- \\
            & 6\,767.770  & 1.830  & $-$2.140  & W14   &  82  & --- & --- & --- \\
            & 6\,772.320  & 3.658  & $-$0.970  & R03   &  47  &  14 &  21 &  32 \\\midrule

Sr\,{\sc i} & 4\,607.33  & 0.00  & $+$0.28    & SN96   & 106  & 36  &  73 &  85 \\\midrule

Y\,{\sc ii} & 4\,883.68  & 1.08  & $+$0.07    & SN96   &  101 & 83  & 108 & 127 \\
            & 5\,087.43  & 1.08  & $-$0.17    & SN96   &  83  &  73 &  91 & 105 \\
            & 5\,200.41  & 0.99  & $-$0.57    & SN96   &  71  &  61 &  80 &  92 \\
            & 5\,205.72  & 1.03  & $-$0.34    & SN96   &  --- & 70  &  87 &  94 \\
            & 5\,289.81  & 1.03  & $-$1.85    & R04    &  22  &  13 &  24 &  38 \\
            & 5\,402.78  & 1.84  & $-$0.44    & R03    &  40  &  27 &  53 &  61 \\\midrule

Zr\,{\sc ii} & 4\,050.32  & 0.71  & $-$1.06   & Lj06   &  50  &  39 & 59  &  69 \\
             & 4\,208.99  & 0.71  & $-$0.51   & Lj06   &  70  &  71 & 77  &  93 \\
             & 4\,210.61  & 1.66  & $-$1.19   & Lj06   & ---  & --- & 23  &  37 \\
             & 4\,317.32  & 0.71  & $-$1.46   & Lj06   & ---  & 29  & 55  &  66 \\
             & 4\,613.95  & 0.97  & $-$1.54   & Lj06   & ---  & --- & 42  & --- \\
             & 5\,112.27  & 1.66  & $-$0.85   & Lj06   & 37   &  17 & 44  &  60 \\
             & 5\,350.35  & 1.77  & $-$1.16   & Lj06   & 23   & --- & 30  &  45 \\\midrule
             
La\,{\sc ii} & 5\,303.53  & 0.32  & $-$1.35   & L01a   & 20  & 6   & 26 & 42  \\ 
             & 5\,805.77  & 0.13  & $-$1.56   & L01a   & 24  & 8   & 29 & 42  \\ 
             & 6\,262.29  & 0.40  & $-$1.22   & L01a   & 31  & 13  & 42 & 76  \\ 
             & 6\,320.43  & 0.17  & $-$1.52   & SN96   & 19  & --- & 29 & 44  \\ 
             & 6\,774.33  & 0.13  & $-$1.71   & VWR    & 14  & --- & 26 & 41  \\\midrule

Ce\,{\sc ii} & 4\,418.79  & 0.86  & $+$0.27 & L09  & ---  & 24   & 58  & --- \\
             & 4\,483.90  & 0.86  & $+$0.10 & L09  & ---  & ---  & 50  & --- \\
             & 4\,486.91  & 0.29  & $-$0.18 & L09  &  49  &  29  & 55  &  66 \\
             & 4\,539.74  & 0.33  & $-$0.08 & L09  & ---  & ---  & 64  & --- \\
             & 4\,560.96  & 0.48  & $-$0.26 & L09  & ---  & ---  & 44  & --- \\
             & 4\,562.37  & 0.48  & $+$0.21 & L09  &  52  &  44  & 69  &  77 \\
             & 5\,187.45  & 1.21  & $+$0.17 & L09  &  27  &  13  & 39  & --- \\
             & 5\,274.24  & 1.04  & $+$0.13 & L09  &  30  &  19  & 43  &  53 \\
             & 5\,330.58  & 0.87  & $-$0.40 & L09  &  16  &  12  & 30  &  40 \\ 
             & 5\,393.39  & 1.10  & $-$0.06 & L09  & 26   &  --- & 36  & --- \\
             & 5\,975.82  & 1.33  & $-$0.45 & L09  & --   & ---  & 19  &  25 \\
             & 6\,043.37  & 1.21  & $-$0.48 & L09  & 10   &  --- & 19  &  27 \\\midrule

Nd\,{\sc ii} & 4\,706.54  & 0.00  & $-$0.71 & DH   & ---  & 27   &  50 & --- \\
             & 4\,763.62  & 0.38  & $-$1.27 & DH   & 10   & ---  & --- & --- \\
             & 4\,777.72  & 0.38  & $-$1.22 & DH   & ---  & ---  &  23 & --- \\
             & 4\,797.15  & 0.56  & $-$0.69 & DH   & ---  & 10   & --- & --- \\
             & 4\,825.48  & 0.18  & $-$0.42 & DH   & ---  & 28   & --- & --- \\
             & 4\,902.04  & 0.06  & $-$1.34 & DH   & ---  & ---  &  25 & --- \\
             & 4\,914.38  & 0.38  & $-$0.70 & DH   & 26   & ---  &  39 &  46 \\
             & 5\,063.72  & 0.98  & $-$0.62 & DH   & ---  & ---  &  19 &  32 \\
             & 5\,092.80  & 0.38  & $-$0.61 & DH   &  28  &  12  &  36 &  50 \\
             & 5\,212.36  & 0.20  & $-$0.96 & DH   & ---  & 11   &  33 &  47 \\
             & 5\,249.58  & 0.98  & $+$0.20 & DH   &  34  &  31  & --- & --- \\
             & 5\,255.51  & 0.20  & $-$0.67 & DH   & ---  & ---  & --- &  55 \\
             & 5\,293.16  & 0.82  & $+$0.10 & DH   & ---  & 22   &  51 &  63 \\
             & 5\,306.46  & 0.86  & $-$0.97 & DH   & ---  & ---  & --- &  20 \\
             & 5\,311.46  & 0.98  & $-$0.42 & DH   & 19   & ---  & --- &  38 \\
             & 5\,319.81  & 0.55  & $-$0.14 & DH   & 39   & ---  &  55 &  65 \\
             & 5\,356.97  & 1.26  & $-$0.28 & DH   & ---  & ---  &  21 & --- \\
             & 5\,371.93  & 1.41  & $-$0.00 & DH   & ---  & ---  &  32 & --- \\    
             & 5\,485.70  & 1.26  & $-$0.12 & DH   & ---  & ---  &  28 & --- \\
             & 5\,740.88  & 1.16  & $-$0.53 & DH   & 10   &  --- &  15 &  23 \\
             & 5\,811.57  & 0.86  & $-$0.86 & DH   & ---  & ---  &  14 & --- \\\midrule 

Sm\,{\sc ii} & 4\,318.94  &  0.28 & $-$0.25 & L06  & --- & ---  & --- &  47 \\
             & 4\,360.71  &  0.25 & $-$0.87 & L06  & --- & 16   & --- & --- \\
             & 4\,362.02  &  0.48 & $-$0.47 & L06  & --- & 19   & --- & --- \\
             & 4\,467.34  &  0.66 & $+$0.15 & L06  & 34  &  --- &  42 &  50 \\
             & 4\,499.48  &  0.25 & $-$0.87 & L06  & 15  &  --- &  19 &  32 \\
             & 4\,523.91  &  0.43 & $-$0.39 & L06  & --- & ---  & --- &  36 \\
             & 4\,566.20  &  0.33 & $-$0.59 & L06  & 22  &  --- &  22 &  39 \\
             & 4\,676.90  &  0.04 & $-$0.87 & L06  & --- & ---  &  23 & --- \\
             & 4\,704.40  &  0.00 & $-$0.86 & L06  & 24  &  --- &  29 &  39 \\
             & 4\,791.60  &  0.10 & $-$1.44 & L06  & --- & ---  &  10 &  14 \\
\end{longtable}

{\sc Notes--}References: (C2003) \citet{chen2003}; (Ca2007) \citet{carretta2007}; (D2002) \citet{depagne2002}; (DH) \citet{denhartog2003}; (DS91) \citet{drake1991}; (E93) \citet{edvardsson1993}; (J2006) \citet{johnson2006}; (L01) \citet{lawler2001a}; (L06) \citet{lawler2006}; (L09) \citet{lawler2009}; (L13) \citet{lawler2013}; (L82) \citet{lambert1982}; (Lj06) \citet{ljung2006}; (MR94) \citet{mcwilliam1994}; (N96) \citet{norris1996}; (PS) \citet{preston2001}; (R03) \citet{reddy2003}; (R04) \citet{reyniers2004}; (R99) \citet{reddy1999}; (S07) \citet{sobeck2007}; (S86) \citet{smith1986b}; (SN96) \citet{sneden1996}; (T02) \citet{takeda2002}; (TH05) \citet{takeda2005}; (TS97) \citet{takeda1997}; (VWR) \citet{vanwinckel2000}; (W14) \citet{wood2014}; (WSM) \citet{wiese1969}.




\end{document}